\documentclass[preprint,12pt]{elsarticle}

\usepackage{amssymb,amsmath,breqn,soul,color,todonotes,booktabs,siunitx}
\usepackage{tikz}
\usetikzlibrary{positioning}
\usetikzlibrary{calc}
\usepackage{pgfplots}
\pgfplotsset{compat=newest}
\usepackage[percent]{overpic}
\usepackage[export]{adjustbox}

\newcommand{\bfA}{\boldsymbol{A}}%
\newcommand{\bfD}{\boldsymbol{D}}%
\newcommand{\bff}{\boldsymbol{f}}%
\newcommand{\bfh}{\boldsymbol{h}}%
\newcommand{\bfI}{\boldsymbol{I}}%
\newcommand{\bfn}{\boldsymbol{n}}%
\newcommand{\bfu}{\boldsymbol{u}}%
\newcommand{\bfv}{\boldsymbol{v}}%
\newcommand{\bfw}{\boldsymbol{w}}%
\newcommand{\bfx}{\boldsymbol{x}}\newcommand{\bfX}{\boldsymbol{X}}%
\newcommand{\bfsigma}{\boldsymbol{\sigma}}%
\newcommand{\bftau}{\boldsymbol{\tau}}%
\newcommand{\bfxi}{\boldsymbol{\xi}}%
\newcommand{\bfepsilon}{\boldsymbol{\epsilon}}%
\newcommand{\bfzero}{\boldsymbol{0}}%

\journal{Journal of the Mechanics and Physics of Solids}

\begin{document}

\begin{frontmatter}

\title{A Darcy--Cahn--Hilliard model of multiphase fluid-driven fracture}

\author[inst1]{Alexandre Gu\'evel}
\author[inst2]{Yue Meng}
\author[inst3]{Christian Peco}
\author[inst4]{Ruben Juanes}
\author[inst1]{John E.\ Dolbow}

\affiliation[inst1]{organization={Department of Mechanical Engineering and Materials, Duke University},
	city={Durham},
	postcode={27708}, 
	state={NC},
	country={USA}}

 \affiliation[inst2]{organization={Department of Geosciences, Princeton University},
	city={Princeton},
	postcode={08544}, 
	state={NJ},
	country={USA}}

\affiliation[inst3]{organization={Department of Engineering Science and Mechanics, Penn State University},
	city={State College},
	postcode={16801}, 
	state={PA},
	country={USA}}

\affiliation[inst4]{organization={Department of Civil and Environmental Engineering, Massachusetts Institute of Technology},
	city={Cambridge},
	postcode={02139}, 
	state={MA},
	country={USA}}

\begin{abstract}
A Darcy--Cahn--Hilliard model coupled with damage is developed to describe multiphase-flow and fluid-driven fracturing in porous media.  The model is motivated by recent experimental observations in Hele--Shaw cells of the fluid-driven fracturing of a synthetic porous medium with tunable fracture resistance. The model is derived from continuum thermodynamics and employs several simplifying assumptions, such as linear poroelasticity and viscous-dominated flow. Two distinct phase fields are used to regularize the interface between an invading and a defending fluid, as well as the ensuing damage. The damage model is a cohesive version of a phase-field model for fracture, in which model parameters allow for control over both nucleation and crack growth. Model-based simulations with finite elements are then performed to calibrate the model against recent experimental results.  In particular, an experimentally-inferred phase diagram differentiating two flow regimes of porous invasion and fracturing is recovered.  
Finally, the model is employed to explore the parameter space beyond experimental capabilities, giving rise to the construction of an expanded phase diagram that suggests a new flow regime.  
\end{abstract}


\end{frontmatter}

\section{Introduction}

The past few decades have witnessed an increased focus on obtaining a better understanding of multiphase flow in deformable porous media.  This interest is driven by the prevalence of such phenomena in both natural and engineered systems, including enhanced geothermal energy, geological carbon storage, and geo-hazards \cite{Juanes2020}. Models that account for various aspects of the process can be extremely useful, as bench-scale experiments that mimic a wide range of subsurface conditions have proven to be elusive.  From a modeling standpoint, this class of problems is nonetheless challenging, with difficulties stemming from a three-way coupling between poromechanics, fracture mechanics, and multiphase fluid flow.  This work introduces a novel double phase-field model in which both the fluid-fluid interface and the fracture surfaces are regularized and distinct. This permits the use of the model to explore a wide range of processes, including hydraulic fracturing, viscous fingering, and their combination. The model is developed within a simplified poroelastic setting in order to limit the number of undetermined parameters and keep the model's response physically tractable. This paves the way for direct comparison with recent experiments examining multiphase-flow driven fracturing \cite{Meng2020,Meng2022,Meng2023}. Beyond its ability to explain these recent experimental observations, the broad potential of the model is utilized to build phase diagrams indicating both existing and new flow regimes.

The current model builds on a considerable amount of previous theoretical developments in which aspects of the underlying physics were examined in isolation or with some degree of coupling.  This includes work on multiphase flow in rigid porous media, regularized models of fracture, and coupled poromechanics and fracture.  In what follows, important theoretical works and recent modeling approaches that informed the current approach are discussed.

To begin with, the topic of multiphase flow in rigid porous media is one with a rich history.   Here, emphasis is placed on the seminal work of Saffman and Taylor~\cite{Saffman1958}, who studied the fluid-fluid displacement instability that occurs when an invading fluid is less viscous than the defending fluid.  This phenomena has since become known as the viscous fingering instability.  Subsequently, 
Lenormand~\cite{Lenormand1986} identified some of the conditions for viscous fingering to occur, in addition to stable invasion and capillary fingering.  More recently, Zhao et al.~\cite{Zhao2016} have augmented the parameter space of consideration by taking into account the role of wettability in multiphase flow.  For a discussion of other important efforts in modeling multiphase flow through rigid porous media, the reader is referred to the recent paper by Juanes et al.~\cite{Juanes2020}.  

The modeling of multiphase flow often gravitates around a Cahn--Hilliard~\cite{Cahn1958} description of the interface between the fluids. Such a regularization circumvents a shortcoming of sharp-interface approaches when it comes to modeling large interface deformation, and, in particular, interface instabilities. The associated general description of multiphase flow led to the Navier--Stokes--Cahn--Hilliard model, introduced by Lowengrub and Truskinovsky~\cite{Lowengrub1998}, which led to the Hele--Shaw--Cahn--Hilliard model~\cite{Lee2002} in the case of viscous flow in Hele--Shaw cells. The latter was proven to asymptotically converge to the corresponding Hele--Shaw sharp interface problem. The Navier--Stokes--Cahn--Hilliard model was later formally derived from a continuum thermodynamic framework by Abels et al.~\cite{Abels2012}. For that, they took inspiration from Gurtin et al.'s work~\cite{Gurtin1996a}, formally deriving from the laws of thermodynamics the previously ad-hoc Navier--Stokes--Cahn--Hilliard models. In Lowengrub and Truskinovsky's derivation~\cite{Lowengrub1998}, the binary fluid velocity was mass-averaged and not divergence-free, whereas in Abels et al.'s derivation~\cite{Abels2012}, the velocity was volume-averaged and, advantageously, divergence-free. Dede et al.\cite{Dede2018} obtained the Hele--Shaw counterpart to the model of Abels et al.~\cite{Abels2012}, along with the proof of sharp interface convergence. Importantly, Papatzacos~\cite{Papatzacos2002}, Cueto-Felgueroso and Juanes~\cite{Cueto2008,Cueto2009}, and others, have applied such Cahn--Hilliard approaches to  multiphase flow through porous media.

The potential for porous media to fracture complicates modeling efforts considerably due to the coupling between fluid flow and fracturing process.  In essence, fluid pressure can drive the formation of new fracture surfaces, and the fracture openings in turn provide preferential paths for fluid flow.  Most of the existing modeling efforts in this area have focused on hydraulic fracture and have been limited to instances in which the media is saturated with only one fluid, as opposed to multiphase flow. We refer the reader to the paper of Chen et al.~\cite{Chen2021} for a recent review of models for hydraulic fracture.  

More recently, advances in fracture mechanics have driven new developments in hydraulic fracture modeling.  We focus in particular on models based on the variational treatment of fracture and accompanying phase-field regularizations \cite{Francfort1998,Bourdin2000}.
Early efforts to extend the basic phase-field for fracture approach to coupling with fluid flow and hydraulic fracture were spearheaded by Miehe and collaborators \cite{Miehe2015,Miehe2016}. 
Their approach was based on Biot's~\cite{Biot1941} and Coussy's~\cite{Coussy2004} poromechanics framework cast in a variational formulation. This model opened avenues for embedding the well-established phase-field fracture approach into poromechanics. In particular, it proposed modifying the permeability to account for the flow enhancement due to fracturing. 
 An alternative approach was derived from the Theory of Porous Media \cite{Bowen1980,deBoer2012,Ehlers2002}, which is the counterpart of Biot's empirical theory \cite{Biot1941} embedded in continuum thermodynamics and the theory of mixtures \cite{Truesdell1960,Bowen1976}. We refer in particular to the models of Ehlers and Luo \cite{Ehlers2017} and Wilson and Landis \cite{Wilson2016}.  The main advances consist of the use of a thermodynamically-consistent framework for the model derivation and constitutive assumptions that permit a transition from Darcy-type flow in the pores to Stokes-type flow within open fractures.  
 For further references on phase-field models of hydraulic fracture, the reader is referred to a recent review by Heider~\cite{Heider2021}.

The full three-way coupling of multiphase flow, poromechanics, and fracture has been investigated to a lesser extent, with some notable recent advances.  
For instance, Holtzman et al.~\cite{Holtzman2012} experimentally identified, for non-cohesive granular media, the fracturing, capillary fingering, and viscous fingering regimes in a phase diagram representing fracturing number versus capillary number.
On the modeling side, Lee et al.~\cite{Lee2018} introduced a hydraulic fracturing model for two-phase flows using fracture phase-field modeling and lubrication theory. In a similar vein, Heider and Sun~\cite{Heider2020} recently proposed a multiphase hydraulic fracturing model but based on the Theory of Porous Media, for both capillary and viscous flow. Importantly, neither the work of Lee et al.~\cite{Lee2018} nor Heider and Sun~\cite{Heider2020} regularized the fluid-fluid interfaces.  As a result, these models are limited in their ability to account for aspects of the interfacial fluid dynamics.  

The closest analog to the model proposed in this work is that of Carrillo and Bourg~\cite{Carrillo2021}, who adopted a fluid mechanics approach.  In particular, they treated soft porous media as an intermediate state between solid and fluid, and which was effectively modeled through visco-plastic rheology.
The model of  Carrillo and Bourg~\cite{Carrillo2021} allows for a clear distinction between the pore and continuum scales due to volume-averaging upscaling, as well as between viscous and capillary effects. While the distinction between viscous and capillary effects was made in the non-fracturing regime in the experimental study of Holtzman et al.~\cite{Holtzman2012}, the work of Carrillo and Bourg~\cite{Carrillo2021} focused on this distinction in the fracturing regime. The latter also identified a regime of non-invasive fracturing, where invasion only occurs in the cracks and not in the porous medium. Importantly, the model of Carrillo and Bourg~\cite{Carrillo2021} relies on a local yield stress criterion to govern fracture and is limited in its ability to examine the potentially important role of fracture mechanics in governing the response of the system. Such a local, stress-based criterion for fracture is also likely to suffer from the kinds of spurious strain localization effects that plague local damage models (see, e.g., \ Pijaudier-Cabot and Bazant~\cite{Pijaudier1987}). Finally, we note that much like the aforementioned three-way coupling models, the fluid-fluid interfaces are not regularized in the work of Carrillo and Bourg~\cite{Carrillo2021}.

Lastly, attention is called to the recent work of Paulin et al.~\cite{Paulin2022} who developed a model for non-cohesive soft porous media which accounts for regularized damage and finite-deformation kinematics. The novelty of this work lies in envisioning the solid, liquid, and gas as three distinct phases, whose interactions are governed by interfacial energies of Cahn--Hilliard type. Although the model was developed for non-cohesive media, it employed a phase-field 
for fracture regularization that prevents the transmission of traction across fully damaged surfaces.  Distinct from many of the aforementioned works, the model of Paulin et al.\cite{Paulin2022} does not insist that damage be irreversible, such that fracture healing can occur.  
 As the model was developed within the framework of continuum thermodynamics, it is thermodynamically consistent.  

The approach described in this manuscript incorporates aspects that are similar to some of the aforementioned literature, with several important distinctions that are motivated by an interest in explaining recent experiments of fluid-driven fracturing of cohesive media \cite{Meng2023}. The configuration consists of a monolayer of beads that are lightly cemented together and placed into a Hele--Shaw cell, and saturated with a viscous defending fluid. The injection of a far less viscous invading fluid into this system gives rise to a clear coupling between fracturing and fluid invasion in which the fracture resistance of the skeleton plays an important role in delineating various regimes of the response. In the configuration examined in this work, one can expect new features such as a non-trivial injection pressure curve and the possibility of viscous fingering instabilities. The former feature can be explained by the fact that both the solid's cement and the defending fluid's viscosity act as a resistance to fracture and invasion of the injected fluid.

Accordingly, for the fracture model, this work adopts a cohesive variation on the well-established phase-field regularization for fracture. In particular, the model described in \cite{Geelen2019} is used.  Importantly, it incorporates an energetic threshold for damage that is insensitive to the choice of regularization length. To accommodate for fluid-fluid interface instabilities, a second phase field is employed, representing the invading fluid saturation. This allows merging the two-pronged effort described above, and, in particular, expanding the space of potential flow regimes. The aforementioned works have identified six different types of flow regimes: uniform invasion, capillary and viscous fingering, invasive capillary and viscous fracturing, and non-invasive fracturing. A supplementary decomposition could be made between cohesive and non-cohesive fracturing. In this contribution, a new regime where viscous fracturing and viscous fingering are combined is revealed. For that, and in line with the experiments, attention is confined to viscous flows (large capillary numbers), leaving aside the capillary regimes. 

This manuscript is divided into two main sections, a theoretical derivation followed by a numerical study in relation to the aforementioned Hele--Shaw experiments.
The modeling strategy is first described, yielding a Darcy--Cahn--Hilliard model of fluid-driven fracture with multiphase flow. The composite model is assembled through a continuum thermodynamic derivation to ensure a consistent coupling of the different components. Namely, these components are inspired from Gurtin et al.~\cite{Gurtin1996a} and Cogswell and Szulczewski~\cite{Cogswell2017} for the multiphase flow, Borja~\cite{Borja2006} for poromechanics, and Ehlers and Luo \cite{Ehlers2017} for hydraulic fracturing. The novelty in the fracture modeling will be to employ a cohesive fracture model \cite{Geelen2019,Hu2020,Hu2021} which allows for controlling both the onset of cracks and their propagation, but also accounts for rate effects through a damage viscosity.
The resulting system of coupled partial differential equations is discretized using the finite-element method. Model-based simulations are then employed to calibrate various parameters against the experimental observations, and to reproduce a phase diagram representing modified capillary number versus cement volume ratio. Finally, the parameter space to build a new phase diagram that extends beyond the experimental observations is explored, yielding a new dimensionless group of interest and a new type of flow regime mixing fracturing and viscous fingering.

\section{Derivation}
\label{sec:section1}

\subsection{Overall framework and approach}
\begin{figure}[h!]
	\centering
	\includegraphics[scale=0.5]{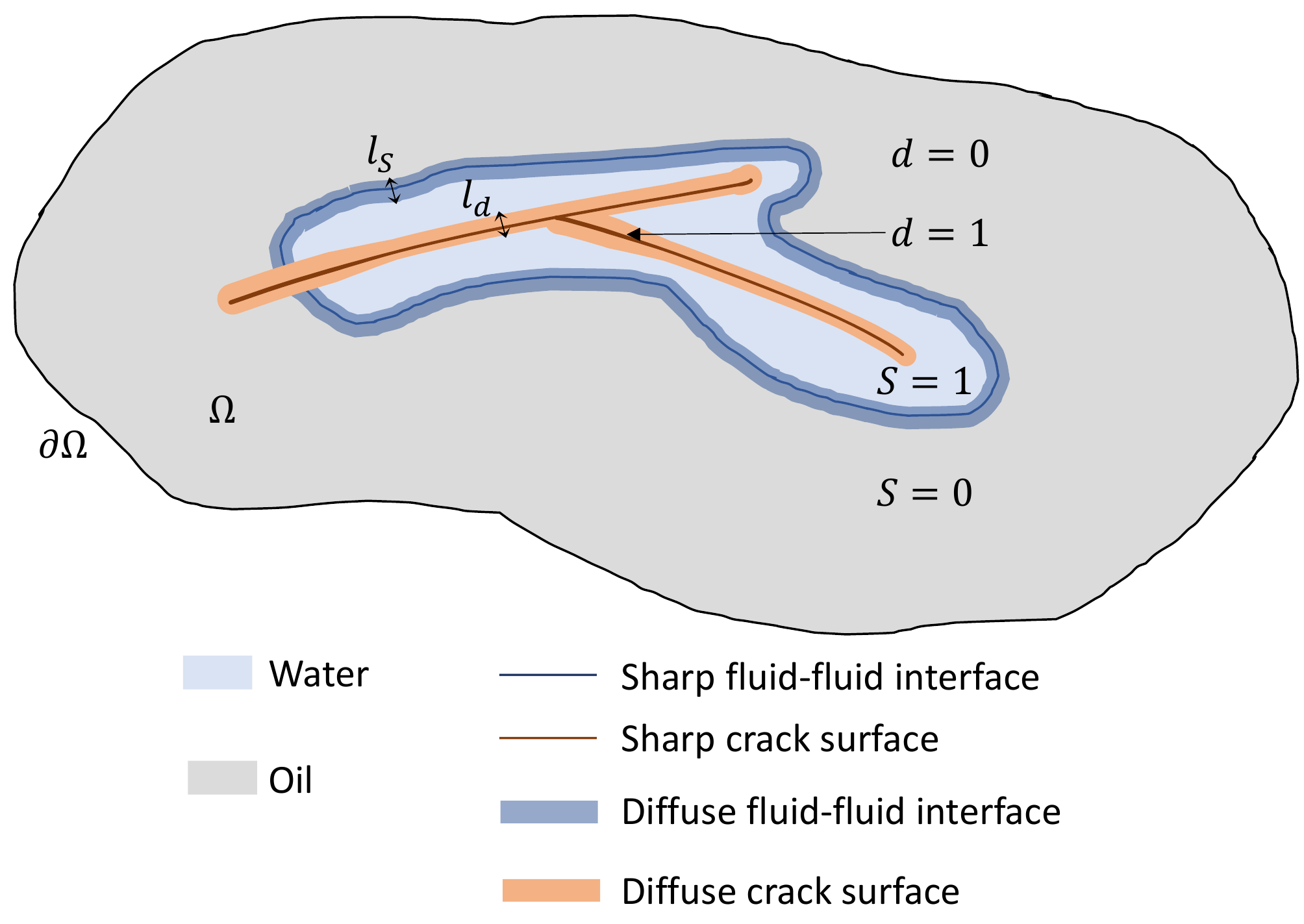}
	\caption[]{Schematic of the modeling of the fluid-fluid interface and crack surfaces through the phase fields $S$ (water saturation) and $d$ (damage), respectively.}
	\label{fig:initial_schema}
\end{figure}

The main novelty in the model lies in the use of two distinct phase fields to regularize geometric features of interest (see Figure~\ref{fig:initial_schema}). The first phase field, denoted by $S$, corresponds to the invading fluid saturation, so that $S=1$ in the invading fluid and $S=0$ in the defending fluid. Intermediate values of $S$ correspond to the diffuse interface, of thickness $l_S$, which regularizes the otherwise sharp interface and ensures a smooth transition from 0 to 1. The second phase field $d$ can be similarly described in that it discriminates between fully cracked regions, where $d=1$, and fully intact regions, where $d=0$. The thickness of the damage band is characterized by the regularization length $l_d$.

Although the resulting phase-field equations will end up looking similar, the saturation phase-field stems from a gradient energy regularization, following the pioneering work of Cahn~\cite{Cahn1958}, whereas the damage phase-field theory stems from a variational regularization of Griffith's crack surface energy, as introduced by Francfort and Marigo~\cite{Francfort1998}. Both phase-field approaches are reconciled by the configurational theory of Fried and Gurtin~\cite{Fried1994,Gurtin1996b,DaSilva2013}, whereby phase-field equations are derived from a fundamental law, the micro-force balance. The latter approach is followed in this work to derive the evolution equations governing both phase fields.

\subsection{Notation and terminology}
Consider a porous medium consisting of three components: one solid and two immiscible pore fluids.  In keeping with Coussy's theory and terminology \cite{Coussy2004}, a porous medium is modeled as the superimposition of a solid \textit{skeleton} continuum (formed by the solid \textit{matrix} and the connected pore space emptied of fluid) and of a pore-filling (binary) fluid continuum. Both subscripts and superscripts will be used to denote fields associated with one of the components.  In particular, $s$ is used to denote the solid component and $\alpha=\{1,2\}$ is used to denote the fluid components, with  $\alpha = 1$ representing the invading fluid and $\alpha = 2$ the defending fluid. 

The volume fractions of the components are denoted by $\phi^s$, $\phi^1$, and $\phi^2$, and satisfy $\phi^s+\phi^1+\phi^2=1$. Similarly, the respective partial densities (at the macroscale) are denoted by $\rho^s$ and $\rho^\alpha$, and the partial Cauchy stresses by $\bfsigma^s$ and $\bfsigma^\alpha$. The corresponding intrinsic quantities (at the mesoscale) are denoted with subscripts. The volume fraction of each fluid in the pores is described by the saturations $S^\alpha$, whereby $S^1+S^2=1$. 

Given the above notation, some key relationships can be summarized as:
\begin{equation}
\begin{aligned}
	\phi^s &= 1-\phi, & \phi^\alpha&=\phi S^\alpha, & \phi^f&=\phi^1+\phi^2=\phi, \\
    \rho^s&= (1-\phi)\rho_s, & \rho^\alpha &= \phi S^\alpha \rho_\alpha, &  \rho^f &= \rho^1+\rho^2,  \\
    \bfsigma^s&=(1-\phi)\bfsigma_s, & \bfsigma^\alpha &= \phi S^\alpha\bfsigma_\alpha, &   \bfsigma^f &=  \bfsigma^1+ \bfsigma^2,
\end{aligned}
\end{equation}
where tensorial quantities are written in bold font. 

\subsection{Main assumptions}
Consider the isothermal motion of a binary fluid through a deformable porous skeleton, and assume that the binary fluid is incompressible, as well as its two individual components, 1 and 2. For simplicity, further assume that the  difference in (constant) densities $\rho_1$ and $\rho_2$ is negligible.
The reader is referred to Abels et al.~\cite{Abels2012} for a derivation when this assumption does not hold. 
Following Gurtin et al.~\cite{Gurtin1996a}, also assume that ``the momenta and kinetic energies of the constituents are negligible when computed relative to the gross motion of the fluid.'' This means in particular that the binary fluid is tracked in time through the material time derivative associated with the average velocity $\bfv_f=S\bfv_1+(1-S)\bfv_2$. Furthermore, assume that the capillary number is sufficiently high such that the pore pressure $p$ is identical in both phases 1 and 2, and also that the flow is sufficiently slow to neglect the fluid's kinetic energy. In other words, consider a viscous flow. 

With regard to the porous skeleton, small deformations and deformation gradients are assumed, 
so that the formulation falls in the framework of linear poroelasticity. 
Unlike the fluid components, the solid matrix is assumed to be compressible, so that its density is not constant.  The resulting continuum is approached through the Theory of Porous Media (see, e.g., \ Ehlers and Luo \cite{Ehlers2017} and references therein), that is, it stems from the superimposition of the continua formed by the different solid and fluid components.

\subsection{Mass balance}
To derive the mass balance equations for the three phases, the derivation here  follows Coussy~\cite{Coussy2004} and Borja~\cite{Borja2006}, combined with Gurtin et al.~\cite{Gurtin1996a}.

The partial mass fluxes associated with the densities $\rho^1$ and $\rho^2$ are denoted by $\tilde\bfh^1$ and $\tilde\bfh^2$, respectively.  As these are internal to the fluid mixture, they satisfy $\tilde\bfh^1+\tilde\bfh^2=\bfzero$. In what follows, a superposed dot is used to denote the material time derivative following the solid skeleton motion.  The solid velocity $\bfv_s$ is therefore 
given by $\bfv_s=\dot{\bfu}$, where $\bfu$ denotes the solid displacement.

With these definitions in hand, the balances for the solid skeleton and the two fluid phases read
\begin{equation}
\label{eq:mass_balances}
\begin{aligned}
    &\dot\rho^s + \rho^s \nabla\cdot \bfv_s =0, \\
    &\dot\rho^\alpha + \rho^\alpha \nabla\cdot \bfv_s + \nabla\cdot\rho_\alpha \bfw_\alpha = -\nabla\cdot  \tilde\bfh^\alpha, 
\end{aligned}
\end{equation}
where $\nabla=\partial/\partial\bfx$ denotes the spatial gradient, and $\bfw_\alpha=\phi^\alpha (\bfv_\alpha-\bfv_s)$ is the Darcy velocity of phase $\alpha$ with respect to the solid skeleton.     

Following Borja's formulation~\cite{Borja2006},  the  compressibility of the solid is assumed to depend only on the pore pressure $p$. More precisely, at the matrix scale (mesoscale), the intrinsic density $\rho_s$ depends only on $p$, and at the skeleton scale (macroscale), the partial density $(1-\phi)\rho_s=\rho^s$ depends only on the partial pressure $(1-\phi)p$. The functional relationships for the matrix and the skeleton follow as, respectively:
\begin{equation}
\label{eq:functional_relationships_Borja}
	\dot{\rho_s} = \frac{\rho_s}{K_s} \dot{p} , \quad (1-\phi)\dot{p} = -K\nabla\cdot\bfv_s,
\end{equation}
where $K_s=\rho_s p'(\rho_s)$ is the matrix bulk modulus {(i.e.\ the bulk modulus of the grains in the experiments studied in this work) and $K$ is the skeleton (drained) bulk modulus, and $p'$ denotes the derivative of $p$. 
The solid mass balance \eqref{eq:mass_balances}$_1$ can then be written in the equivalent following forms:
\begin{equation}
\begin{aligned}
	\label{eq:equivalent_solid_mass_balance}
		 &\dot{\overline{1-\phi}} + (1-\phi)\left(\nabla\cdot\bfv_s + \frac{\dot{p}}{K_s}\right)=0, \\
		 &\dot{\overline{1-\phi}}+(b-\phi)\nabla\cdot \bfv_s=0,
\end{aligned}
\end{equation}
where $b=1-K/K_s$ denotes Biot's coefficient. 

As for the mass balances of the fluid constituents, dividing \eqref{eq:mass_balances}$_{2}$ by the constant densities yields
\begin{equation}
\begin{aligned}
    &\dot{\overline{\phi S_1}} + \phi S_1 \nabla\cdot \bfv_s + \nabla\cdot  \bfw_1 = -\nabla\cdot \bfh^1, \\
    &\dot{\overline{\phi S_2}} + \phi S_2 \nabla\cdot \bfv_s + \nabla\cdot  \bfw_2 = -\nabla\cdot \bfh^2,
\end{aligned}
\end{equation}
where $\bfh^\alpha=\tilde\bfh^\alpha/\rho_\alpha$. Summing up the two previous equations, the mass balance of the binary fluid can be written as
\begin{equation}
\label{eq:mass_balance_binary_fluid}
    \dot\phi + \phi \nabla\cdot \bfv_s + \nabla\cdot  \bfw =0,
\end{equation}
where $\bfw = \bfw_1+\bfw_2 = \phi\bfv_{f/s}$ is the Darcy velocity of the binary fluid and $\bfv_{f/s}=S(\bfv_1-\bfv_s)+(1-S)(\bfv_2-\bfv_s)$, the velocity of the binary fluid with respect to the solid, is defined here for later use. Then, summing up the mass balances of the solid \eqref{eq:equivalent_solid_mass_balance}$_2$ and of the binary fluid \eqref{eq:mass_balance_binary_fluid} yields the mass balance of the mixture:
\begin{equation}
\label{eq:mass_balance_mixture}
    b\nabla\cdot\bfv_s + \nabla\cdot\bfw=0.
\end{equation}

Finally, for later use, it will be helpful to have at hand some alternative expressions of the mass balances.  A superposed circle (e.g.,\ $\overset{\circ}a$), is used to denote the material time derivative following the binary fluid.
The mass balances of each fluid component and of the resulting binary fluid can thus be rewritten as
\begin{equation}
\begin{aligned}
   &\overset{\circ}{\overline{\phi S_\alpha}}  = -\nabla\cdot\bfh^\alpha, \\
    &\overset{\circ}{\phi} = 0,
\end{aligned}
\end{equation}
where the incompressiblity of each fluid component ($\nabla\cdot \bfv_\alpha=0$) and of the binary fluid ($\nabla\cdot (S \bfv_1 +(1-S) \bfv_2)=0$) have been used.

\subsection{Force balances}
The macro-force balances of the solid and binary fluid read
\begin{equation}
\label{eq:macro-force_balance}
\begin{aligned}
    &\nabla\cdot\bfsigma^s + \bff^{fs}=0, \\
    &\nabla\cdot\bfsigma^f + \bff^{sf}=0,
\end{aligned}
\end{equation}
where $\bfsigma^s$ and $\bfsigma^f$ denote the partial solid and fluid Cauchy stresses, respectively, while $\bff^{fs}$ and $\bff^{sf}$ are the forces exerted by the fluid upon the solid and vice-versa. The effects of external body forces like gravity have been neglected. The forces satisfy the reciprocity relationship $\bff^{fs}+\bff^{sf}=0$. They are also referred to as  momentum productions in Ehlers and Luo's work~\cite{Ehlers2017}. They result from the superimposition of the continua formed by the solid and the binary fluid. 

The above form of the force balance for the binary fluid stems from neglecting the relative momenta of the constituents, as in Gurtin et al.'s derivation~\cite{Gurtin1996a}. In particular, recall that the pore pressure is assumed to be the same in both phases. As a result, the force balance for the mixture, described by the total stress $\bfsigma=\bfsigma^s+\bfsigma^f$ reads
\begin{equation}
\label{eq:total_force_balance}
    \nabla\cdot\bfsigma=0.
\end{equation}
A derivation similar to that in \cite{Gurtin1996a} would also show that the symmetry of $\bfsigma^s$ and $\bfsigma^f$ follows from the corresponding balances of angular momentum. 
The fluid stress tensor can be written in terms of its shear component $\bftau$ and the pore pressure $p$ as  $\bfsigma^f=\phi(\bftau-p \bfI)$.   It bears emphasis that, unlike in Gurtin et al.'s derivation~\cite{Gurtin1996a}, due to the poromechanical coupling, the pore pressure $p$ is not indeterminate.

Finally, following the configurational theory of Fried and Gurtin~\cite{Fried1994,Gurtin1996b}, the phase fields $S$ and $d$ are governed by the following microforce balances:
\begin{equation}
\label{eq:micro-force_balances}
\begin{aligned}
    &\nabla\cdot\bfxi_d + \pi_d=0, \\
    &\nabla\cdot\bfxi_S + \pi_S=0,
\end{aligned}
\end{equation}
where $\bfxi$ denotes the microstress that is energetically conjugate to the gradient of the respective phase field, and $\pi$ the microforce that is conjugate to the phase field.

\subsection{Dissipation inequality}

Within the framework of continuum thermodynamics,  constitutive restrictions for the governing equations are now invoked. The partial Helmholtz free energies per unit volume for the solid and binary fluid are denoted by $\psi^s$ and  $\psi^f$, respectively. The chemical potential of constituent $\alpha$ associated with the flux $\bfh^\alpha$ is denoted by $\mu_\alpha$. Since the mass fluxes are complementary, in this section attention is restricted to constituent 1 and  $\bfh$ is used in place of $\bfh^1$ and $\mu$ is used to denote the difference in chemical potentials, i.e. $\mu=\mu_1-\mu_2$. Similarly, $S$ is used in place of $S_1$. 

The second law expressed in the form of the dissipation inequality requires that, for a given control volume $\Omega_0$ in the reference configuration, the rate at which the energy of $\Omega_0$ increases (following Coussy's formulation~\cite{Coussy2004} and neglecting the kinetic energy),
\begin{dmath}
	\label{eq:energy}
	\int\limits_{\Omega_0}{\left((\dot\psi^s+\overset{\circ}{\psi^f}\right)} \ \text{d}V,
\end{dmath}
is limited by the external work on $\Omega_0$,
 \begin{dmath}
 	\int\limits_{\partial \Omega_0}{\bfsigma^s \bfn \cdot \dot{\bfu}} \ \text{d}A + \int\limits_{\partial \Omega_0}{\dot{d}\bfxi_d\cdot \bfn} \ \text{d}A + \int\limits_{\partial \Omega_0}{\bfsigma^f \bfn \cdot \bfv_f} \ \text{d}A + \int\limits_{\partial \Omega_0}{\overset{\circ}{S}\bfxi_S\cdot \bfn} \ \text{d}A,
 \end{dmath}
 plus the rate at which energy is transported to $\Omega_0$ by diffusion~\cite{Gurtin1996a},
  \begin{dmath}
   - \int\limits_{\partial \Omega_0}{\mu_\alpha \bfh^\alpha \cdot n} \ \text{d}A= - \int\limits_{\partial \Omega_0}{\mu\bfh\cdot n} \ \text{d}A.
  \end{dmath}

The local form of the dissipation inequality expressed above is then given by
\begin{dmath}
\label{eq:dissipation_inequality_local_1}
  \dot\psi^s+\overset{\circ}{\psi^f} \le {\bfsigma^s:\nabla{\bfv_s}} + { \bfv_s\cdot\nabla\cdot\bfsigma^s } + {\bfxi_d\cdot\nabla\dot{d}} + {\dot{d}\nabla\cdot\bfxi_d} + {\bfsigma^f:\nabla\bfv_f} + {\bfv_f\cdot\nabla\cdot\bfsigma^f} + \bfxi_S\cdot\nabla{\overset{\circ}{S}} + \overset{\circ}{S}\nabla\cdot\bfxi_S - \mu\nabla\cdot \bfh - \bfh\cdot\nabla\mu,
\end{dmath}
where the linear poroelasticity approximation $\partial/\partial\bfX\approx\partial/\partial\bfx$ has been invoked to retain the spatial gradients. Using the force balances and the mass balance of the invading fluid, and noting in particular that $\overset{\circ}{\phi}=0$, one obtains
\begin{dmath}
\label{eq:dissipation_inequality_local_2}
    \dot\psi^s+\overset{\circ}{\psi^f} \le {\bfsigma^s:\nabla{v_s}} - {\bfv_s\cdot\bff^{fs}} + {\bfxi_d\cdot\nabla\dot{d}} - {\pi_d\dot{d}} + {\bfsigma^f:\nabla\bfv_f} - {\bfv_f\cdot\bff^{sf}} + \bfxi_S\cdot\nabla{\overset{\circ}{S}} - \pi_S{\overset{\circ}{S}} + \mu\phi \overset{\circ}{S} - \bfh\cdot\nabla\mu.
\end{dmath}
The familiar poromechanical concept of total effective stress can be obtained by adding the null term $(\bfsigma-\bfsigma^s-\bfsigma^f):\nabla\bfv_s$ to \eqref{eq:dissipation_inequality_local_2}, yielding
\begin{dmath}
\label{eq:dissipation_inequality_local_3}
   \dot\psi^s + \overset{\circ}{\psi^f} \le {\bfsigma:\nabla\bfv_s} + {\bfxi_d\cdot\nabla\dot{d}} - {\pi_d\dot{d}} + {\bfsigma^f:\nabla\bfv_{f/s}} - {\bfv_{f/s}\cdot\bff^{sf}} + \bfxi_S\cdot\nabla{\overset{\circ}{S}} + (\phi\mu - \pi_S){\overset{\circ}{S}} - \bfh\cdot\nabla\mu.
\end{dmath}
Following the derivation by Ehlers and Luo~\cite{Ehlers2017}, the pore pressure is then employed as a Lagrange multiplier, multiplying \eqref{eq:mass_balance_mixture} by $p$ and adding it to the previous inequality to obtain:
\begin{dmath}
	\label{eq:dissipation_inequality_local_4}
	\dot\psi^s+\overset{\circ}{\psi^f} \le {\bfsigma':\nabla\bfv_s} + {\bfxi_d\cdot\nabla\dot{d}} - {\pi_d\dot{d}} + {\phi\bftau:\nabla\bfv_{f/s}} + {p\bfv_{f/s}\cdot\nabla\phi} - {\bfv_{f/s}\cdot\bff^{sf}}  + \bfxi_S\cdot\nabla\overset{\circ}{S} + (\phi\mu - \pi_S){\overset{\circ}{S}} - \bfh\cdot\nabla\mu,
\end{dmath}
where $\bfsigma'=\bfsigma+b p \bfI$ denotes the total effective stress. The last step consists in identifying the energy-conjugates, that is, the products of forces and rates of state variables. For that, one must carefully commute gradients and time derivatives, especially when the former are spatial quantities whereas the latter are material quantities, as here, in which case one must employ the following commutator identity (see e.g.\ \cite{Gurtin1996a}):
\begin{dmath}
	\nabla\left(\frac{\text{d}^\alpha\varphi}{\text{d}t}\right)= \frac{\text{d}^\alpha}{\text{d}t}\left(\nabla\varphi\right) + (\nabla\bfv_\alpha)^T\nabla\varphi.
\end{dmath}
Thereupon, 
\begin{equation}
	\begin{aligned}
	    &\nabla\dot{d} = \dot{\overline{\nabla{d}}} + (\nabla\bfv_s)^T\nabla{d} \approx  \dot{\overline{\nabla{d}}}, \\
		&\nabla\dot{\bfu} = \dot{\overline{\nabla\bfu}}+(\nabla\bfv_s)^T\nabla\bfu \approx \dot{\overline{\nabla\bfu}}, \\
	    &\nabla\overset{\circ}{S} = \overset{\circ}{\overline{\nabla{S}}} + (\nabla\bfv_f)^T\nabla{S},
	\end{aligned}
\end{equation}
where the approximations in the two first equations stem from the linear poroelasticity assumption that second-order terms in deformation are negligible (see e.g.\ section 3 in MacMinn et al.~\cite{Macminn2016}). In particular, assume that $d$ varies like $\phi$ since their evolution is tightly coupled, as the model-based simulations in this work will indicate.  The local dissipation inequality \eqref{eq:dissipation_inequality_local_4} then reads:
\begin{dmath}
	\label{eq:dissipation_inequality_local_5}
	\dot\psi^s+\overset{\circ}{\psi^f} \le {(\bfsigma'+\nabla{S}\otimes\bfxi_S):\dot{\overline{\nabla\bfu}}} + \bfxi_d\cdot\dot{\overline{\nabla{d}}} - {\pi_d\dot{d}} + {(\phi\bftau+\nabla{S}\otimes\bfxi_S):\nabla\bfv_{f/s}} +
	 {(p\nabla\phi -\bff^{sf})\cdot\bfv_{f/s}}  + \bfxi_S\cdot\overset{\circ}{\overline{\nabla{S}}}  + (\phi\mu - \pi_S)\overset{\circ}{S} - \bfh\cdot\nabla\mu.
\end{dmath}
The change of variable $\hat\mu=\phi\mu$ is applied and the hat notation is dropped thereafter for clarity. Identifying as state variables $\{\nabla\bfu,d,\nabla{d},\dot{d},\bfv_{f/s},\bfD,S,\nabla{S},\mu,\nabla\mu\}$, where $\bfD=(\nabla\bfv_{f/s}+{\nabla\bfv_{f/s}}^T)/2$ denotes the strain rate tensor for the binary fluid~\cite{Gurtin1996a}, and invoking Ehlers' principle of phase separation~\cite{Ehlers2017}, consider the constitutive dependencies:
\begin{equation}
	\begin{aligned}
&{\psi^s=\hat{\psi^s}(\nabla\bfu,d,\nabla{d},\dot{d})=\hat{\psi^s}(\dots)_s}, \\ &{\psi^f=\hat{\psi^f}(\bfv_{f/s},\bfD,S,\nabla{S},\mu,\nabla\mu)=\hat{\psi^f}(\dots)_f}, \\ &{\bfsigma'=\hat\bfsigma'(\dots)_s}, \\ &{\bfxi_d=\hat\bfxi_d(\dots)_s}, \\ &{\pi_d=\hat\pi_d(\dots)_s}, \\ &{\bftau=\hat\bftau(\dots)_f}, \\ &{\bfxi_S=\hat\bfxi_S(\dots)_f}, \\ &{\pi_S=\hat\pi_S(\dots)_f}, \\ &{\bftau=\hat\bftau(\dots)_f}, \\ &{p=\hat p(\dots)_f}, \\ &{\bfh=\hat\bfh(\dots)_f}.
	\end{aligned}
\end{equation}
This hereby-defined constitutive model satisfies the dissipation inequality \eqref{eq:dissipation_inequality_local_5} if and only if
\begin{dmath}
	\label{eq:dissipation_inequality_local_6}
	 {\left(\hat\bfsigma'(\dots)_s+\nabla{S}\otimes\hat\bfxi_S(\dots)_f - \frac{\partial\hat\psi^s(\dots)_s}{\partial{\nabla\bfu}}\right):\dot{\overline{\nabla\bfu}}} + {\left(\hat\bfxi_d(\dots)_s - \frac{\partial\hat\psi^s(\dots)_s}{\partial{\nabla{d}}}\right)\cdot\dot{\overline{\nabla{d}}}}  - {\left(\hat\pi_d(\dots)_s + \frac{\partial\hat\psi^s(\dots)_s}{\partial{d}}\right) \dot{d}} - \frac{\partial\psi^s}{\partial\dot{d}}\ddot{d} + {\left(\phi\hat\bftau(\dots)_f+\nabla{S}\otimes\hat\bfxi_S(\dots)_f \right):\bfD} + {\mathcal{A}\left(\nabla{S}\otimes\hat\bfxi_S(\dots)_f \right):\bfA} +
	{\left(\hat{p}(\dots)_f\nabla\phi -\bff^{sf}\right)\cdot\bfv_{f/s}} - \frac{\partial\psi^f}{\partial\bfv_{f/s}}\overset{\circ}{\bfv_{f/s}} - \frac{\partial\psi^f}{\partial\bfD}\overset{\circ}{\bfD} +
	{\left(\hat\bfxi_S(\dots)_f - \frac{\partial\hat\psi^f(\dots)_f}{\partial{\nabla{S}}}\right)\cdot\overset{\circ}{\overline{\nabla{S}}}} + \left(\mu- \hat\pi_S(\dots)_f - \frac{\partial\hat\psi^f(\dots)_f}{\partial{S}}\right)\overset{\circ}{S} - \hat\bfh(\dots)_f\cdot\nabla\mu - \frac{\partial\psi^f}{\partial\mu}\overset{\circ}{\mu} - \frac{\partial\psi^f}{\partial\nabla\mu}\overset{\circ}{\overline{\nabla\mu}} \ge 0,
\end{dmath}
where  the tensor $\nabla\bfv_{f/s}$ has been decomposed into its symmetric $\bfD$ and antisymmetric part $\bfA$, and the symmetry of $\bftau$ has been employed. 

Now, since $\dot{\overline{\nabla\bfu}}$, $\dot{\overline{\nabla{d}}}$, $\ddot{d}$, $\bfA$, $\overset{\circ}{\bfv_{f/s}}$, $\overset{\circ}\bfD$, $\overset{\circ}{\overline{\nabla{S}}}$, $\overset{\circ}{S}$, $\overset{\circ}\mu$ and $\overset{\circ}{\overline{\nabla\mu}} $ appear linearly in \eqref{eq:dissipation_inequality_local_6}, employing the Coleman--Noll procedure~\cite{Coleman1963} yields the following constitutive restrictions:
\begin{equation}
\begin{aligned}
	\label{eq:constitutive_restrictions}
		&\frac{\partial\psi^s}{\partial\dot{d}} = 0, \quad \frac{\partial\psi^f}{\partial\bfv_{f/s}}=0, \quad \frac{\partial\psi^f}{\partial\bfD}=0, \quad \frac{\partial\psi^f}{\partial\mu}=0, \quad \frac{\partial\psi^f}{\partial\nabla\mu}=0, \\
		&\mathcal{A}\left(\nabla{S}\otimes\hat\bfxi_S(\dots)_f\right) =0, \\
		&\hat\bfsigma'(\dots)_s+\nabla{S}\otimes\hat\bfxi_S(\dots)_f = \frac{\partial\hat\psi^s(\dots)_s}{\partial{\nabla\bfu}}, \\
		&\hat\bfxi_d(\dots)_s = \frac{\partial\hat\psi^s(\dots)_s}{\partial{\nabla{d}}}, \\ 
		&\hat\bfxi_S(\dots)_f = \frac{\partial\hat\psi^f(\dots)_f}{\partial{\nabla{S}}}, \\ 
		&\hat\pi_S(\dots)_f = \mu + \frac{\partial\hat\psi^f(\dots)_f}{\partial{S}},
\end{aligned}
\end{equation}
leaving the following reduced dissipation inequality:
\begin{dmath}
	\label{eq:dissipation_inequality_local_7}
     -{\left(\hat\pi_d(\dots)_s + \frac{\partial\hat\psi^s(\dots)_s}{\partial{d}}\right) \dot{d}} + {\left(\phi\hat\bftau(\dots)_f+\nabla{S}\otimes\hat\bfxi_S(\dots)_f \right):\bfD} +
    {\left(\hat{p}(\dots)_s\nabla\phi -\bff^{sf}\right)\cdot\bfv_{f/s}} - \hat\bfh(\dots)_f\cdot\nabla\mu \ge 0.
\end{dmath}
From \eqref{eq:constitutive_restrictions}$_2$, one can deduce that $\nabla{S}\otimes\bfxi_S=\bfxi_S\otimes\nabla{S}$, which implies that $\left(\nabla{S}\otimes\bfxi_S\right)\bfxi_S=\left(\bfxi_S\otimes\nabla{S}\right)\bfxi_S$, i.e.\ $\left(\nabla{S}\cdot\bfxi_S\right)\bfxi_S = \left(\bfxi_S\cdot\bfxi_S\right)\nabla{S}$: the vectors $\bfxi_S$ and $\nabla{S}$ are collinear. The corresponding collinearity coefficient $\kappa$ is introduced and chosen constant for simplicity, so that 
\begin{equation}
	\label{eq:Korteweg_stress}
	 \bfxi_S = \kappa\nabla{S} = \frac{\partial\psi^f}{\partial\nabla{S}}.
\end{equation}
The reasoning above is similar to that invoked by Abels et al.~\cite{Abels2012}. Furthermore, Gurtin's invariance argument~\cite{Gurtin1996b} implies that the constitutive functions can depend on $\nabla\bfu$ only through the strain tensor of the solid $\bfepsilon=(\nabla\bfu+\nabla\bfu^T)/2$, so that $\nabla\bfu$ can be replaced here above by $\bfepsilon$. Finally, a sufficient condition to satisfy \eqref{eq:dissipation_inequality_local_7} is to require that each term be non-negative:
\begin{equation}
\begin{aligned}
	\label{eq:solution_dissipation_inequality}
		&\beta\dot{d} = \nabla\cdot\frac{\partial\psi^s}{\partial{\nabla{d}}} - \frac{\partial\psi^s}{\partial{d}} , \\
		&\phi\bftau+\kappa\nabla{S}\otimes\nabla{S} = C_{St}\bfD, \\
		&p\nabla\phi-\bff^{sf} = C_{Da}\bfv_{f/s}, \\
		&\bfh = -M\nabla\mu,
\end{aligned}
\end{equation}
where $\beta\ge0$ is the damage viscosity, $C_{St}\ge0$ is a coefficient proportional to the binary fluid viscosity coefficient (in keeping with Newtonian fluids' rheological law~\cite{Abels2012,Gurtin1996a}), $C_{Da}\ge0$ is a coefficient inversely proportional to the binary fluid mobility, and $M\ge0$ is the Cahn-Hilliard mobility. These quantities need not be constant, and their dependencies will be specified later.

\subsection{Specification of the free energy}

Following the previous thermodynamic analysis, the solid free energy is of the form $\hat\psi^s(\bfepsilon,d,\nabla{d})$ and is expressed in the following well-established form (see, e.g.,~\cite{Geelen2019,Hu2020,Hu2021}):
\begin{equation}
	\label{eq:solid_free_energy}
	\hat\psi^s(\bfepsilon,d,\nabla{d}) = g(d)\psi_e^A(\bfepsilon) + \psi_e^I(\bfepsilon) + G_c h(d,\nabla{d}),
\end{equation}
where $g(d)$ is a stiffness degradation function, with $g(0)=1$ and $g(1)=0$, and $\psi_e^A$ and $\psi_e^I$ denote the active (tensile) and inactive (compressive) parts of the elastic energy, respectively. The critical fracture energy is denoted by $G_c$, and $h$ is the crack density function defined by
\begin{equation}
	\label{eq:crack_density_function}
	h(d,\nabla{d}) = \frac{1}{c_0 l_d}\left(w(d)+l_d^2\lvert\nabla{d}\rvert^2\right),
\end{equation}
where $c_0>0$ is a normalization constant, $w(d)$ is the local dissipation function~\cite{Marigo2016}, and $l_d$ is the regularization length for the phase field $d$.

The fluid free energy is of the form $\hat\psi^f(S,\nabla{S})$ and is expressed following Cogswell and Szulczewski's formulation~\cite{Cogswell2017} as:
\begin{equation}
\label{eq:fluid_free_energy}
	\hat\psi^f(S,\nabla{S}) = \frac{12\gamma}{l_S}\left(f(S) + \frac{l_S^2}{16} \lvert\nabla{S}\rvert^2\right),
\end{equation}
where $\gamma$ is the surface tension between the two fluid constituents, $l_S$ is the regularization length for the phase field $S$, and $f(S)=S^2(1-S)^2$ is chosen as a double-well potential. Thereby, the gradient coefficient $\kappa$ discovered in the thermodynamic analysis \eqref{eq:Korteweg_stress} is given by $\kappa=3\gamma l_S/2$.

\subsection{Fracture phase-field model}

Different choices of the functions $g(d)$ and $w(d)$ lead to different fracture phase-field models. Here, we adopt a cohesive fracture model~\cite{Geelen2019,Hu2020,Hu2021} so as to control both crack nucleation and propagation. The former is controlled by a nucleation energy $\psi_c$, acting as a damage threshold, and the latter by the critical fracture energy $G_c$. Importantly, such a model allows the regularization length $l_d$ to be prescribed independently of the fracture properties, provided it satisfies an upper bound. Following \cite{Hu2021}, the functions $g(d)$ and $w(d)$ are taken to be
\begin{equation}
\begin{aligned}
	\label{eq:fracture_functions}
	&g(d) = \frac{(1-d)^2}{(1-d)^2+md(1+d)}, \\
	&w(d)=d,
\end{aligned}
\end{equation}
where $m=G_c/c_0l_d\psi_c$ and $c_0=4\int_0^1{\sqrt{w(d)}\text{d}d}=8/3$. To ensure the aforementioned advantageous properties of the cohesive model, the condition $m\ge 1$ should be satisfied~\cite{Geelen2019}, which yields the following upper bound for the regularization length:
\begin{equation}
	\label{eq:fracture_upper_bound}
	l_d \le \frac{3G_c}{8\psi_c}.
\end{equation}
The fracture phase-field evolution equation is then obtained from \eqref{eq:solution_dissipation_inequality}$_1$, \eqref{eq:solid_free_energy} and \eqref{eq:fracture_upper_bound}:
\begin{equation}
\begin{aligned}
	\label{eq:fracture_equation}
	\beta\dot{d} &= \frac{G_c}{c_0 l_d}\left(2l_d^2\triangle{d}-w'(d)\right) -  g'(d,\psi_c)\psi_e^A, \\ &= \frac{3G_c}{8 l_d}\left(2l_d^2\triangle{d}-1\right) -  g'(d,\psi_c) \psi_e^A,
\end{aligned}
\end{equation}
where the dependence of the degradation function $g$ on the the nucleation energy $\psi_c$ has been emphasized.  
The term $\beta\dot{d}$ accounts for the overall rate-dependency of fracture. A source of this rate-dependency is the viscous effect experimentally observed during fracturing (see Figure \ref{fig:comparison-simu-MIT-exp}b), due to the latent cement that slows down the separation of the beads. A second potential source is the frictional contact between  the beads and the two plates of the Hele--Shaw cell. 

To complete the phase-field for the fracture part of the model, the irreversibility condition $\dot{d}\ge 0$ is enforced. Finally, the spectral decomposition of the elastic energy, as described in the work of Miehe et al.~\cite{Miehe2010}, is invoked:
\begin{equation}
	\label{eq:spectral_decomposition}
	\begin{aligned}
	&\psi_e^A(\bfepsilon)= \frac{\lambda_s}{2}\langle\epsilon_v\rangle^2 + \mu_s\sum_{i=1}^{3}\langle\epsilon_i\rangle^2, \\
	&\psi_e^I(\bfepsilon)= -\frac{\lambda_s}{2}\langle-\epsilon_v\rangle^2 - \mu_s\sum_{i=1}^{3}\langle-\epsilon_i\rangle^2,
	\end{aligned}
\end{equation}
where $\langle\cdot\rangle$ denotes the Macaulay brackets and $\epsilon_i$ the principal strains.

\subsection{Binary fluid evolution equations}

In light of the experimental conditions examined in this work, the flow in the cracks is assumed to take the form of a Poiseuille flow between the Hele--Shaw cell plates, which are maintained at a constant separation.  
More complex models for the fluid flow in the cracks can be found in the works of Wilson and Landis~\cite{Wilson2016} or Ehlers and Luo~\cite{Ehlers2017}, for example.
Consistent with the simplifying assumption employed in this work, $C_{St}$ is assumed to vanish and only the terms describing a Darcy flow are maintained, such that  \eqref{eq:solution_dissipation_inequality}$_{2,3}$ can be injected into the fluid force balance \eqref{eq:macro-force_balance}$_2$ to obtain:
\begin{equation}
	C_{Da}\bfv_{f/s}=-\phi\nabla{p} - \nabla\cdot(\kappa\nabla{S}\otimes\nabla{S}).
\end{equation}
The fluid mobilities are noted $\lambda_\alpha$. Upon choosing $C_{Da}=\phi^2/(\lambda_1+\lambda_2)$ and recalling that $\bfw = \phi\bfv_{f/s} = \bfw_1+\bfw_2$, a sufficient condition to describe the flow of the two constituents within the current thermodynamic framework is
\begin{equation}
	\label{eq:fluid_equation}
	\bfw_\alpha = -\lambda_\alpha\left(\nabla{p} + \phi^{-1}\nabla\cdot(\kappa\nabla{S}\otimes\nabla{S})\right),
\end{equation}
which is Darcy's law augmented by a capillary (or Korteweg) stress \cite{Gurtin1996a}. More specifically, the fluid mobilities are given by 
\begin{equation}
	\lambda_\alpha = \frac{k(\phi)k_{r\alpha}(S_\alpha)}{\eta_\alpha},
\end{equation}
where $k$ denotes the absolute permeability, $k_{r\alpha}$ the relative permeability of phase $\alpha$, and $\eta_\alpha$ the dynamic viscosity of phase $\alpha$. The absolute permeability is expressed  as a function of $\phi$ following Kozeny--Carman's formula (see, e.g.,~\cite{Coussy2004}), through
\begin{equation}
	\label{eq:permeability}
	k(\phi)=\frac{\phi^3}{(1-\phi)^2}\bar{k},
\end{equation}
where $\bar{k}$ is the intrinsic permeability. Importantly for the coupling between poromechanics and fracturing, the porosity is expected to reach $\phi=1$ in the cracks, whereby the permeability $k$ is expected to diverge. This potential issue is circumvented by imposing an upper bound $k_{max}=h^2/12$, where $h$ denotes the height of the Hele--Shaw cell, in line with the aforementioned assumption of Poiseuille flow within the crack. Thereby, the combination of Darcy flow in the intact porous skeleton and Poiseuille flow in the cracks is obtained through imposing $k=\text{min}(k(\phi),k_{max})$.

As for the relative permeabilities, the model of Fourar and Lenormand~\cite{Fourar1998} (also used in Cueto--Felgueroso and Juanes's model~\cite{Cueto-felgueroso2014}) derived for viscous coupling in fractures is employed in this work:
\begin{equation}
\begin{aligned}
	\label{eq:relative_permeabilities}
	&k_{r1} = S_1^3 + \frac{3}{2\text{M}}S_1(1-S_1)(1+S_1), \\
	&k_{r2} = \frac{1}{2}S_2^2(3-S_2),
\end{aligned}
\end{equation}
where constituent 1 is chosen to be the non-wetting (invading) phase and  constituent 2 is chosen to be the wetting (defending) phase, and $\text{M}=\eta_2/\eta_1$ is the viscosity ratio.

The Cahn--Hilliard mobility $M$ in \eqref{eq:solution_dissipation_inequality}$_4$ is then specified following Cogswell and Szulczewski's model~\cite{Cogswell2017}:
\begin{equation}
	M=\frac{k}{\overline{\eta}},
\end{equation}
where $\overline{\eta}=2\eta_1\eta_2/(\eta_1+\eta_2)$ is the harmonic average of the viscosities of the two fluid constituents.

\subsection{Model summary}
\label{subsec:Model summary}

To conclude this theoretical section, a summary  of
the evolution equations in the model that will be discretized is provided below \eqref{eq:summary_equations}. Since the two constituents are directly related through $S_1+S_2=1$ and $\bfh^1+\bfh^2=\bfzero$, only the evolution of constituent 1 (the invading fluid) is described; recall the notation $S=S_1$. The model therefore consists of the mass balance for the invading fluid, the total mass balance for the binary fluid, the solid mass balance, the total macro-force balance, and the damage evolution (through the damage micro-force balance), which are respectively given by
\begin{equation}
	\boxed{
\begin{aligned}
	\label{eq:summary_equations}
	&\dot{\overline{\phi S}} + \nabla\cdot\bfw_1(p,\phi) + \phi S \dot\epsilon_v = \nabla\cdot M(\phi)\nabla\mu(S), \\
	&\dot\phi + \nabla\cdot\bfw(p,\phi) + \phi\dot\epsilon_v=0, \\
	&\phi = \phi_0 + (1-\phi_0)\epsilon_v + \frac{1-\phi_0}{K_s}p, \\
	&\nabla\cdot\left(\bfsigma'(\bfepsilon,S) - bp\bfI\right) = 0, \\
	&\beta\dot{d} = \frac{3G_c}{8 l_d}\left(2l_d^2\triangle{d}-1\right) -  g'(d,\psi_c)\psi_e^A, \quad \dot{d}\ge0 .
\end{aligned}
}
\end{equation}
The porosity evolution equation \eqref{eq:summary_equations}$_3$ derives from integrating the solid mass balance \eqref{eq:equivalent_solid_mass_balance}$_1$ (also see equation~(3.34) in Borja~\cite{Borja2006}) and linearizing the resulting exponential function around the reference state  $\{\epsilon_v=0, p/K_s=0\}$. This system of equations \eqref{eq:summary_equations} will be solved for $\mu$, $p$, $\phi$, $\bfu$, and $d$, respectively. It is closed with initial and boundary conditions, and the following system of constitutive equations, given by \eqref{eq:fluid_equation}, \eqref{eq:constitutive_restrictions}$_6$, \eqref{eq:micro-force_balances}$_2$, \eqref{eq:fluid_free_energy}, \eqref{eq:constitutive_restrictions}$_3$, and \eqref{eq:solid_free_energy}, as:
\begin{equation}
\begin{aligned}
	\label{eq:summary_equations_secondary}
	&\bfw_\alpha = -\lambda_\alpha\left(\nabla{p} + \phi^{-1}\nabla\cdot(\kappa\nabla{S}\otimes\nabla{S}) \right) , \\
	&\mu = \frac{12\gamma}{l_S} \left(f'(S) - \frac{l_S^2}{8}\triangle S \right), \\
	&\bfsigma' = g(d)\frac{\partial\psi_e^A}{\partial\bfepsilon}+\frac{\partial\psi_e^I}{\partial\bfepsilon} - \kappa\nabla{S}\otimes\nabla{S}.
\end{aligned}
\end{equation}
It bears emphasis that the third terms in \eqref{eq:summary_equations}$_1$ and \eqref{eq:summary_equations}$_2$ are of second order in terms of the deformation, which may appear at odds with the assumption of linear poroelasticity. Through extensive testing, the retention of these higher order terms was found to significantly improve the calibration of model-based simulations to experimental observations.  In particular, these terms permit the model to better represent the apparent coupling between the evolution of the saturation front and fracture propagation.

\subsection{Dimensionless form}

The governing equations can be rendered dimensionless with the assistance of a characteristic length $l_0$, velocity $v_0$, fluid viscosity $\eta_0=\eta_2$, permeability $k_0$, and surface tension $\gamma_0=\gamma$. It is also useful to define the characteristic pressure $p_0=v_0 \eta_0/l_0$, fluid mobility $\lambda_0=k_0/\eta_0$, Cahn--Hilliard mobility $M_0=k_0/\eta_0=\lambda_0$, and chemical potential $\mu_0=\gamma_0/l_0$. In the following, dimensionless quantities are noted with an asterisk. The previous system \eqref{eq:summary_equations} then becomes
\begin{equation}
\begin{aligned}
	\label{eq:dimensionless_equations_intermed}
	&\frac{v_0}{l_0}\dot{\overline{\phi S}}^* + \frac{v_0}{l_0}\nabla^*\cdot\bfw_1^* + \frac{v_0}{l_0}\phi S_1 \dot\epsilon_v^* = \frac{M_0\gamma_0}{l_0^3}\nabla^*\cdot M^*\nabla^*\mu^*, \\
	&\frac{v_0}{l_0}\dot\phi + \frac{v_0}{l_0}\nabla^*\cdot\bfw^* + \frac{v_0}{l_0}\phi\dot\epsilon_v^*=0, \\
	&\phi = \phi_0 + (1-\phi_0)\epsilon_v + \frac{1-\phi_0}{p_0 K_s^*}p_0 p^*, \\
	&\frac{p_0}{l_0}\nabla\cdot\left({\bfsigma'}^* - bp^*\bfI\right) = 0, \\
	&\frac{v_0\beta}{l_0}\dot{d}^* = \frac{3G_c}{8 l_d}\left(2\frac{l_d^2}{l_0^2}\triangle^*{d}-1\right) -  g'(d,\psi_c)\psi_e^A, \quad \dot{d}^*\ge0.
\end{aligned}
\end{equation}
Similarly, the dimensionless forms of the secondary equations \eqref{eq:summary_equations_secondary} read:
\begin{equation}
\begin{aligned}
	\label{eq:dimensionless_equations_secondary_intermed}
	&\bfw_\alpha^* = -\lambda_\alpha^*\left(\frac{\lambda_0 p_0}{v_0 l_0}\nabla^*{p^*} + \frac{\lambda_0 \kappa}{v_0 l_0^3} \phi^{-1}\nabla\cdot(\nabla{S}\otimes\nabla{S})\right), \\
	&\frac{\gamma_0}{l_0}\mu^* = \frac{12\gamma_0}{l_S} \left(f'(S) - \frac{l_S^2}{8l_0^2}\triangle S \right), \\
	&p_0{\bfsigma'}^* = p_0 g(d)\frac{\partial{\psi_e^A}^*}{\partial\bfepsilon}+p_0\frac{\partial{\psi_e^I}^*}{\partial\bfepsilon} - \frac{\kappa}{l_0^2}\nabla^*{S}\otimes\nabla^*{S}.
\end{aligned}
\end{equation}
The final system of equations in dimensionless form is then given by:
\begin{equation}
\begin{aligned}
	\label{eq:dimensionless_equations}
	&\dot{\overline{\phi S}} + \nabla\cdot\bfw_1 + \phi S_1 \dot\epsilon_v = \text{Pe}^{-1}\nabla\cdot M\nabla\mu, \\
	&\dot\phi + \nabla\cdot\bfw + \phi\dot\epsilon_v=0, \\
	&\phi = \phi_0 + (1-\phi_0)\epsilon_v + \frac{1-\phi_0}{ K_s} p, \\
	&\nabla\cdot\left({\bfsigma'} - bp\bfI\right) = 0, \\
	&\text{B}\dot{d} = 2 \text{L\textsubscript d}^2\triangle{d}-1 -  g'(d,\psi_c)\text{D\textsubscript f}, \quad \dot{d}\ge0,
\end{aligned}
\end{equation}
along with
\begin{equation}
\begin{aligned}
	\label{eq:dimensionless_equations_secondary}
	&\bfw_\alpha = -\lambda_\alpha\text{Da}\left(\nabla{p} + \frac{3}{4}\text{Ca}^{-1}\text{L\textsubscript S} \phi^{-1}\nabla\cdot(\nabla{S}\otimes\nabla{S})\right), \\
	&\mu = \frac{12}{\text{L\textsubscript S}} \left(f'(S) - \frac{\text{L\textsubscript S}^2}{8}\triangle S \right), \\
	&{\bfsigma'} = g(d)\frac{\partial{\psi_e^A}}{\partial\bfepsilon}+\frac{\partial{\psi_e^I}}{\partial\bfepsilon} - \text{K}\nabla{S}\otimes\nabla{S},
\end{aligned}
\end{equation}
where we have dropped the asterisk notation for clarity.  The dimensionless groups employed in the above are given by
\begin{equation}
\begin{aligned}
	\label{eq:dimensionless_groups}
	\text{Pe}&=\frac{v_0l_0^2}{M_0\gamma_0}, & \quad \text{Ca}&=\frac{\eta_0v_0}{\gamma_0}, & \quad \text{Da}&=\frac{k_0}{l_0^2}, \\
	\text{K}&=\frac{\kappa}{p_0l_0^2}, & \text{M}&=\frac{\eta_2}{\eta_1}, & \text{L\textsubscript S}&=\frac{l_S}{l_0}, \\
	\quad \text{B}&=\frac{8l_dv_0\beta}{3l_0G_c}, & \quad \text{D\textsubscript f} &= \frac{8l_d\psi_e^A}{3G_c}, & \text{L\textsubscript d}&=\frac{l_d}{l_0},
\end{aligned}
\end{equation}
denoting the P\'eclet number, the capillary number, the Darcy number, the Korteweg number, the viscosity ratio (appearing indirectly through \eqref{eq:relative_permeabilities}), the normalized saturation regularization length, the damage viscosity number, the fracture driving force, and the normalized damage regularization length, respectively. The two first rows represent  groups associated with fluid fields, whereas the third row represents the groups associated with solid fields. Note that there are only five independent fluid groups since $\text{Pe}=4\text{Da}^{-1}\text{Ca}/3$. The expression for the P\'eclet number matches the one obtained by Abels et al.\cite{Abels2012} for instance. Recall that $\kappa=3\gamma l_S/4$. 

In practice, it is also useful to define a modified capillary number $\text{Ca}^*$ introduced by Holtzman et al.~\cite{Holtzman2012} and employed in the experiments \cite{Meng2023} considered in this work. This number is the ratio of the driving force (i.e., \ the viscous pressure drop $\delta{p}_{vis}$) to the capillary pressure force $\delta{p}_{cap}$. The former can be estimated as
\begin{equation}
\label{eq:viscous_pressure_drop}
    \delta{p}_{vis}\sim\frac{\eta_o w_i l_0}{k_0}\sim\frac{\eta_o q l_0}{h \delta k_0},
\end{equation}
where $w_i$ is the injection Darcy velocity of water, $q$ is the corresponding volumetric flow rate, which scales as $q\sim w_i h\delta$, $h$ is the height of the Hele--Shaw cell, $\delta$ is a characteristic grain size, which will be the diameter of the beads used in the experiments studied in this work, and the characteristic length $l_0$ is taken as the radius $r_o$ of the Hele--Shaw cell. The capillary pressure drop is estimated as
\begin{equation}
    \delta{p}_{cap}\sim\frac{\gamma}{\delta}.
\end{equation}
Thereby, the modified capillary number reads
\begin{equation}
	\label{eq:modified_capillary_number}
	\text{Ca}^* \sim \frac{q\eta_ol_0}{\gamma h k_0} \sim \text{Ca} \frac{l_0 \delta}{k_0}.
\end{equation}

Given the assumption of viscous-dominated flow, $\text{Ca}^{-1}\gg1$, and therefore $\text{Pe}^{-1}\ll1$, but also $\text{K}\ll1$, so that one can anticipate that the terms containing these numbers should not play a significant role. That said, the Cahn--Hilliard term on the right side of \eqref{eq:dimensionless_equations} plays an important interface regularization role, as mentioned by Cogswell and Szulczewski~\cite{Cogswell2017}.

\section{Numerical study}

 In this section, the ability of the model to reproduce the salient aspects of recent experimental observations, and in particular, to recover a phase diagram distinguishing two flow regimes is examined. Model-based simulations are then employed to make predictions for system responses corresponding to regions of parameter space that are beyond recent experimental observations. 

\subsection{Numerical implementation}

The system of equations \eqref{eq:summary_equations}, \eqref{eq:summary_equations_secondary} is discretized through the finite-element method and implemented within the Multiphysics Object Oriented Simulation Environment (MOOSE)~\cite{Permann2020}. For the coupling with phase-field fracture, the RACCOON extension \cite{RACCOON} is used. The resulting system of equations is solved in a staggered fashion, whereby the poromechanical equations \eqref{eq:summary_equations}$_{1-4}$ and the damage equation \eqref{eq:summary_equations}$_5$ are solved iteratively at each time step until convergence is reached. For the damage equation, the irreversibility constraint $\dot{d}\ge0$ is enforced with a primal-dual active set strategy~\cite{Hu2020}. Each of the two sub-systems of nonlinear equations are solved with the Newton--Raphson method. Therein, the rate terms are integrated in time using an implicit backward Euler approximation. Finally, the matrix inversions are obtained through an LU decomposition, using the preconditioner MUMPS (Multifrontal Massively Parallel sparse direct Solver)~\cite{MUMPS} available within the PETSc library. 

\subsection{Description of recent experiments}
\label{subsec:Description of the experimental support}

Recent experiments conducted at MIT concerning the fluid-driven fracturing of a cemented pack of beads in a Hele--Shaw cell \cite{MengLiJuanes-personal-2022,Meng2023} are briefly described in this subsection.  Various detailed observations from these experiments are useful to both calibrate various parameters in the current model and demonstrate some of its capabilities and limitations.  

The experimental configuration consists of a monolayer of spherical beads that are cemented together.
Both the beads and cement are made from polyurethane rubber, which enables the use of photoporoelasticity, a technique developed by Li et al.~\cite{Li2021} to visualize the effective stress field. 
The amount of cement can be controlled, effectively modulating the fracture resistance of the skeleton. 
The skeleton is placed between the two plates of the Hele--Shaw cell, and the plates are held at a constant distance from each other with four outer clamps.  The system is then saturated with oil and water is injected at the center.  As such, the entire system effectively approximates a deformable porous media interacting with multiphase flow.   Aspects of the experimental setup are very similar to those conducted for non-cohesive granular media, as described in \cite{Meng2022}.

Importantly, in the most recent experiments \cite{Meng2023}, the cement renders the granular media cohesive, with the degree of cohesion effectively delineating two regimes in the response of the system.  For sufficiently high cement volume fractions, the solid skeleton deforms but does not fracture as the water invades the media and expels the defending oil.  Conversely, as the amount of cement is decreased, the fracture resistance of the solid skeleton also decreases, and at some point the forces on the skeleton are sufficiently large to break the cement bonds and permit fracture patterns to emerge.  

The values of the main parameters are listed in Table~\ref{tab:table_summary}. Note in particular the oil viscosity $\eta_o$, which is five orders of magnitude larger than the water viscosity $\eta_w$.  This gives rise to a relatively large viscous pressure drop $\delta{p}_{vis}$ (see \eqref{eq:viscous_pressure_drop}), acting as a driving force responsible for fracturing the cemented beads. The high contrast in fluid viscosities was necessary to compensate for the limited injection pressure delivered by the laboratory pump.

Another particularity of the type of experiment  focused on here is the fact that water leaks off the fracture: the invading water flows both in the opening fractures and in the pore spaces of the undamaged domain.  Following the description by Carrillo and Bourg~\cite{Carrillo2021}, this type of response is referred to as ``invasive fracturing". 
A last feature that bears emphasis is that the injection pressure keeps increasing as the cracks form, almost until the water is expelled out of the cell. This is expected to be the case as long as cracks are continuing to grow (see, e.g., \ Figure~7 in \cite{Santillan2018}), which is the case in the experiments and in the corresponding model-based simulations.

 It bears emphasis that the mechanical and fracture response of the cement-bead network was experimentally characterized independently in dry conditions.  The experimentally measured bulk moduli are listed in Table ~\ref{tab:table_summary}.   As for the fracture toughness, the critical mode-I stress intensity factor $K_{1c}$ was measured via a dog-bone tension test. As a function of the cement ratio $C$ (expressed in \%), it was experimentally found that $K_{1c}=\SI{0.15C+0.59}{kPa.\metre^{1/2}}$, so that
\begin{equation}
	\label{eq:Gc}
G_c=\frac{K_{1c}(C)^2}{E/(1-\nu^2)}.
\end{equation}
This assumes brittle failure, which is confirmed by the stress-strain output of the tensile test showing a linear increase of the stress until failure, followed by a sharp decrease; that is, there is negligible plastic deformation and ductile failure can be neglected.

\subsection{Boundary-value problem and initial conditions}

Consider the two-dimensional annular domain defined by an inner and outer radius, as shown in Figure~\ref{fig:BVP}.  The boundary conditions are prescribed on the inner and outer surfaces in a manner meant to approximate the aforementioned experimental setup.  The values of all model parameters contained within the boundary conditions are provided in Table~\ref{tab:table_summary}.

\begin{figure}[h!]
	\centering
	\includegraphics[scale=0.6]{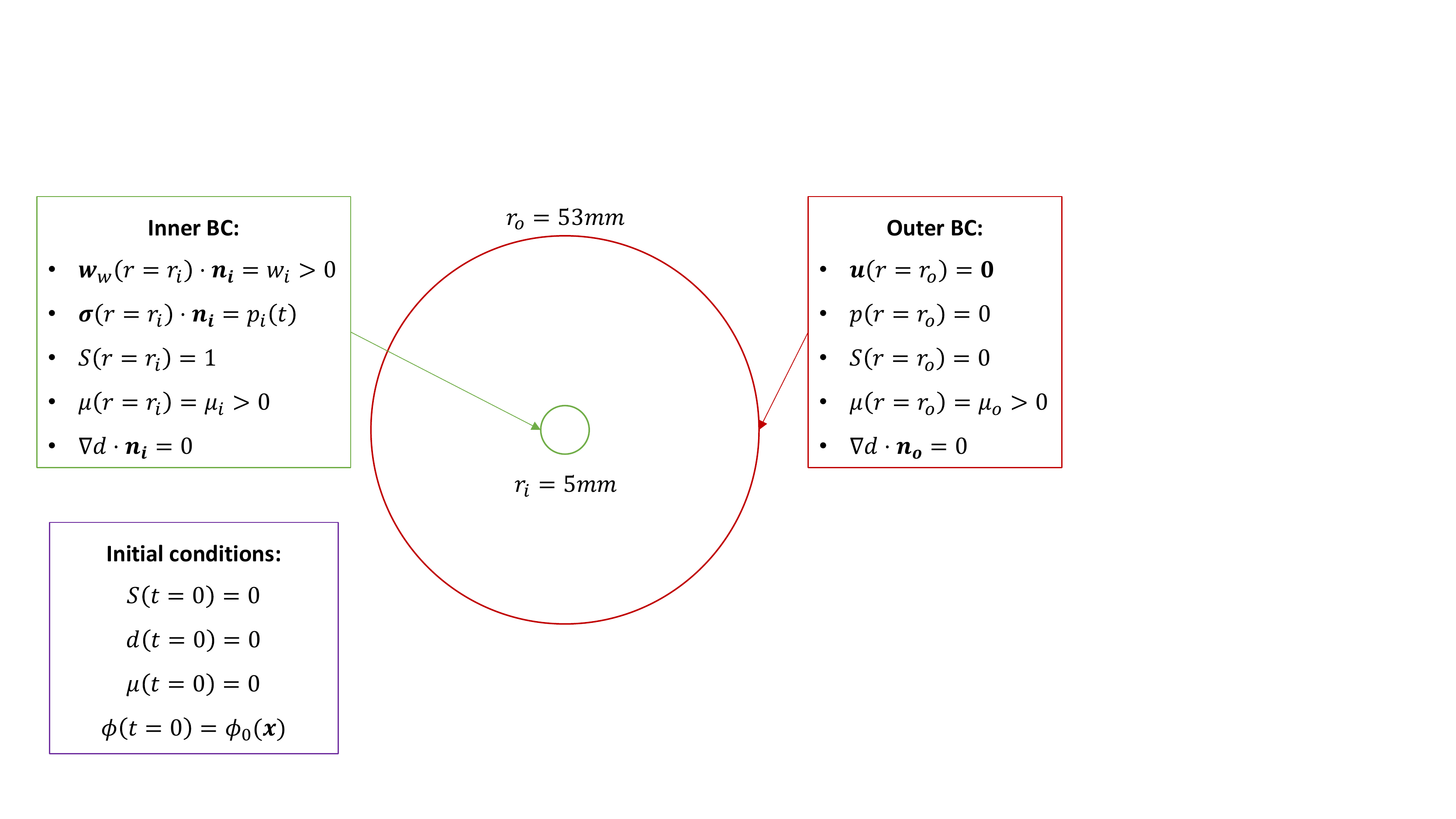}
	\caption[]{Schematic of the boundary-value problem along with boundary and initial conditions. For the boundary conditions, $\bfn_i$ and $\bfn_o$  denote the inward  normal and outward normal at the inner and outer boundaries, respectively.}
	\label{fig:BVP}
\end{figure}

For the mechanical fields,  boundary conditions that approximate fluid injection at a constant flow rate at the inner boundary and constrained outflow conditions at the outer boundary  are prescribed.   In particular, at the inner boundary, the Darcy velocity of the invading fluid is fixed at a value $w_i$ chosen to match the experimental injection pressure curve. This is akin to a traction boundary condition on the displacement field~\cite{Rehbinder1995}, whereby the injection pressure is applied to the total stress. At the outer boundary, the pressure is fixed to 0 (the air pressure in the laboratory, used as a reference pressure) and the displacement field is held fixed, consistently with the mechanical constraint on the cemented bead network in this region.
Finally, zero Neumann boundary conditions on the damage field are imposed at both boundaries, as is standard in phase-field models of fracture.  

 For the saturation field $S$, Dirichlet conditions of $S=1$ and $S=0$ are prescribed at the inner and outer boundaries, respectively.  These conditions reflect the fact that the invading fluid enters at the inner boundary, while the defending fluid exits the domain at the outer boundary.  The latter boundary condition obviously limits the applicability of the model to the point in time when the leading edge of the invading fluid just reaches the outer boundary.

The boundary conditions on the chemical potential are inspired by the work of Dong \cite{Dong2014}.  At the inner boundary, the chemical potential is fixed to a small value $\mu_i > 0$ to facilitate the incoming flow of fluid.  
On the outer boundary, Dong's zero-flux condition is replaced by a Dirichlet boundary condition $\mu_o > 0$.  
In the calibration of the model-based simulations against the experimental observations, the point in time when the injection pressure peaks is observed to be sensitive to $\mu_o$.  Accordingly, the magnitude of $\mu_o$ is adjusted to obtain the best match between the simulated and experimental injection curves.  

The initial conditions, as indicated in Figure~\ref{fig:BVP}, consist in setting all variables to 0, except the porosity field. In this work, the influence of the microstructure stemming from the arrangement of the beads is not examined in detail.   Instead, a continuum perspective and focus on the influence of the hydro-mechanical parameters is adopted. However, some heterogeneity in the initial conditions is required to break the inherent symmetry of the problem and facilitate localization. To effect this, the initial porosity field is assumed to be uniformly random. More precisely, a porosity value, varying between 0.2 and 0.6, is randomly assigned to each mesh element (see Figure~\ref{fig:ini-poro-fields}a). This ensures that the mean porosity value is 0.4, which corresponds to the measured experimental value.  Although the effect of this porosity range is not studied extensively, in limited testing, changes to the upper and lower bounds in the porosity range are found to yield mostly quantitative and not qualitative differences in the model-based simulation results.  

In addition, to establish spatial convergence with the discretization, a different porosity field that does not vary with mesh refinement is considered. To construct such a field, a constant porosity field is used with three small circular regions of high porosity near the inner boundary, giving rise to the initial porosity field shown in Figure~~\ref{fig:ini-poro-fields}b.  

\begin{figure}[h!]
	\centering
	\begin{overpic}[width=0.45\linewidth]{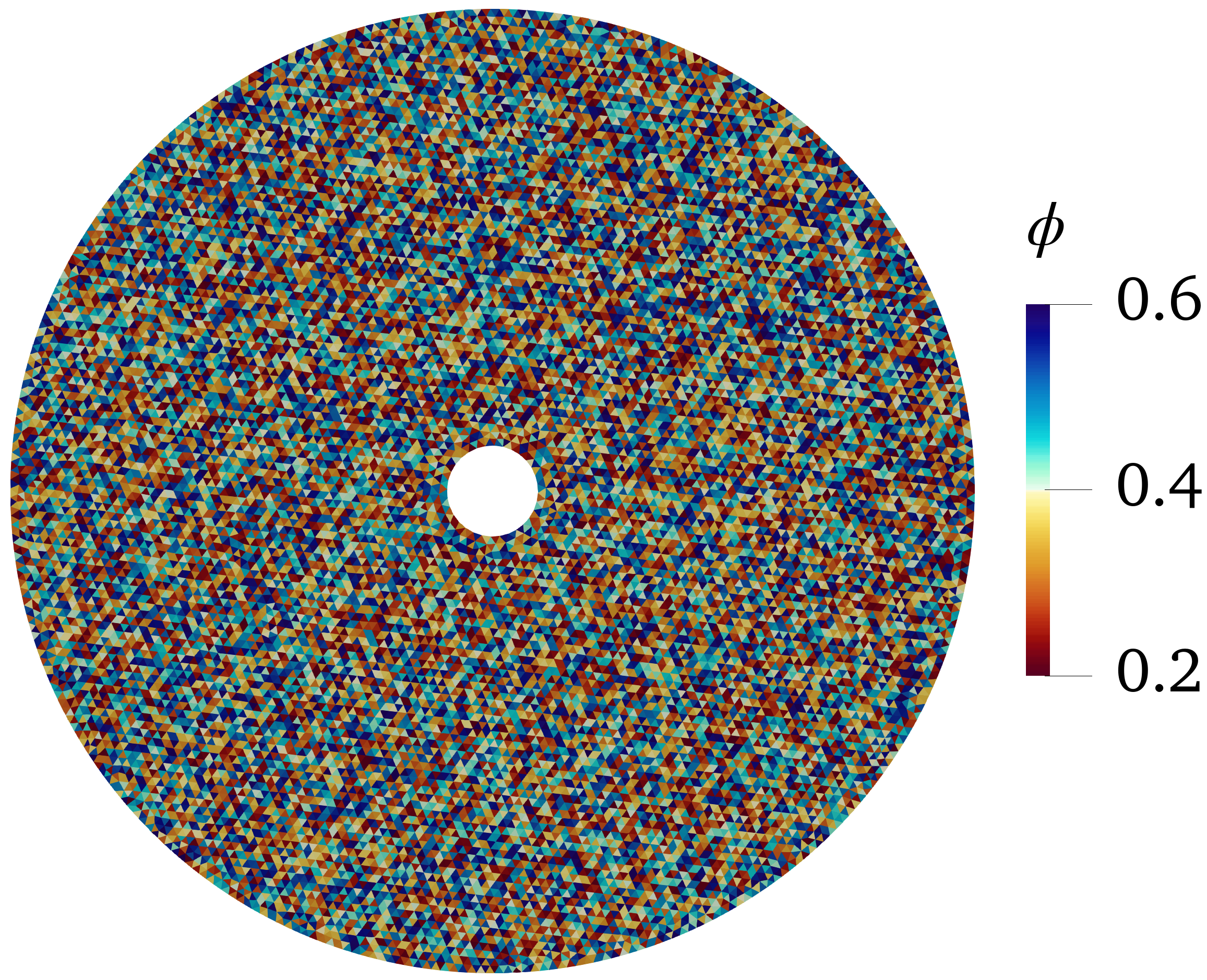}
		\put (-8,65) {\color{black} $a)$}
	\end{overpic}
	\hspace{0.05\linewidth}
	\begin{overpic}[width=0.45\linewidth]{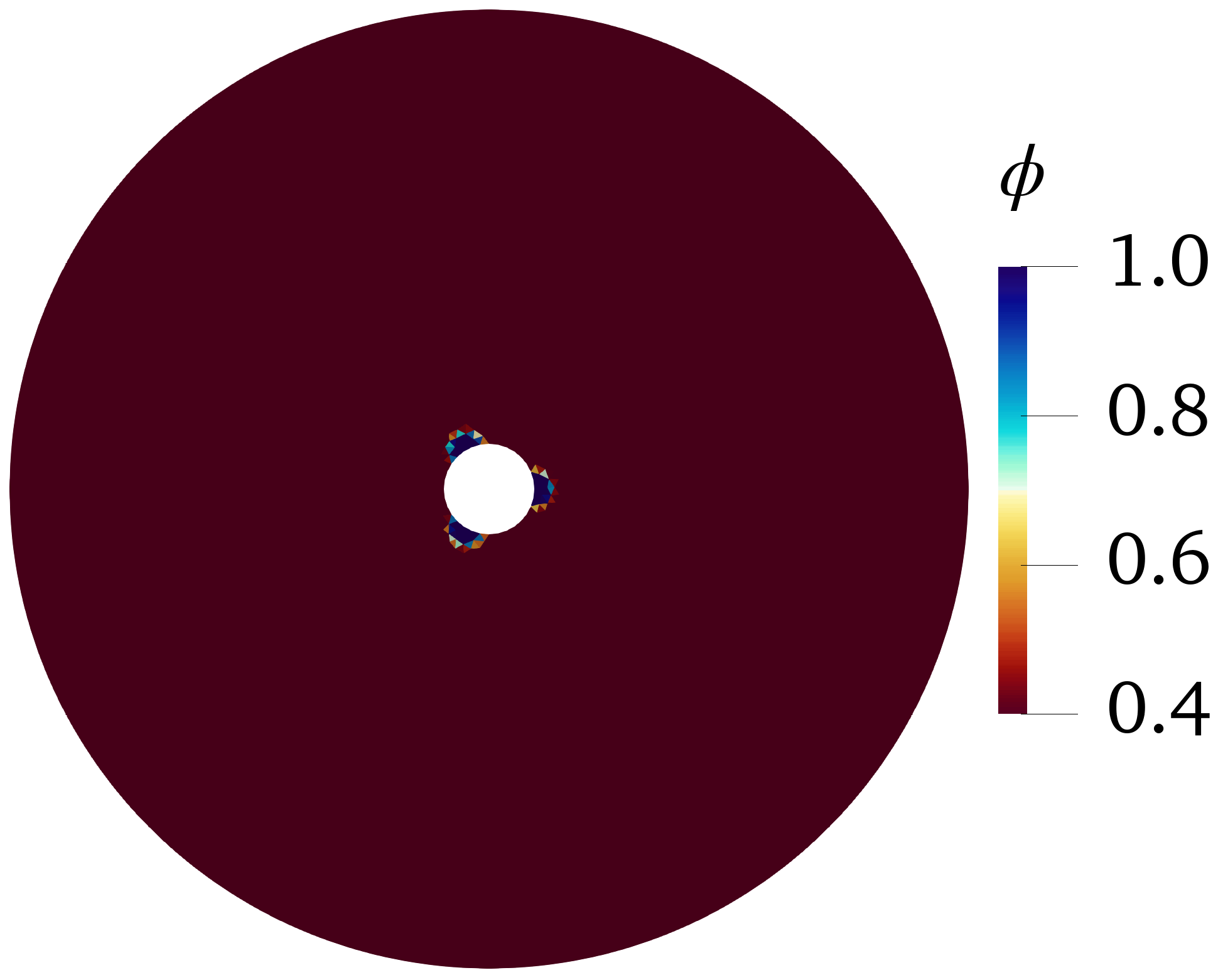}
		\put (-8,65) {\color{black} $b)$}
	\end{overpic}
	\caption{Initial porosity fields used in this work. a) Random porosity field with mean value $\bar{\phi}=0.4$. b) Constant porosity field with existing porosity pockets near the inner boundary employed in the mesh convergence study.}
	\label{fig:ini-poro-fields}
\end{figure}

\subsection{Spatial convergence study}

The ability of the model and accompanying discretization to obtain spatial convergence in selected quantities of interest is first established.  A series of simulations with increasing mesh refinement for the initial boundary-value problem described in Figure~\ref{fig:BVP} using the initial porosity field shown in Figure~\ref{fig:ini-poro-fields}b are performed.  Figure~\ref{fig:mesh-conv} shows plots of the evolution of the injection pressure over time and of the maximum hoop stress for a reference mesh with uniform spacing $h=\SI{1.0}{\milli\metre}$ (20,672 elements) and two coarser meshes. The results indicate that the reference mesh satisfies the requirement of mesh convergence. Accordingly, subsequent studies reported in this manuscript use unstructured meshes of triangular elements with mesh spacing $h=\SI{1}{\milli\metre}$. Importantly, this level of spatial resolution is sufficient to capture the regularization lengths indicated in Table~\ref{tab:table_summary}.  More precisely, the phase field interfaces for $S$ and $d$ span approximately four elements.

\begin{figure}[h!]
	\centering
	\begin{overpic}[width=0.48\linewidth]{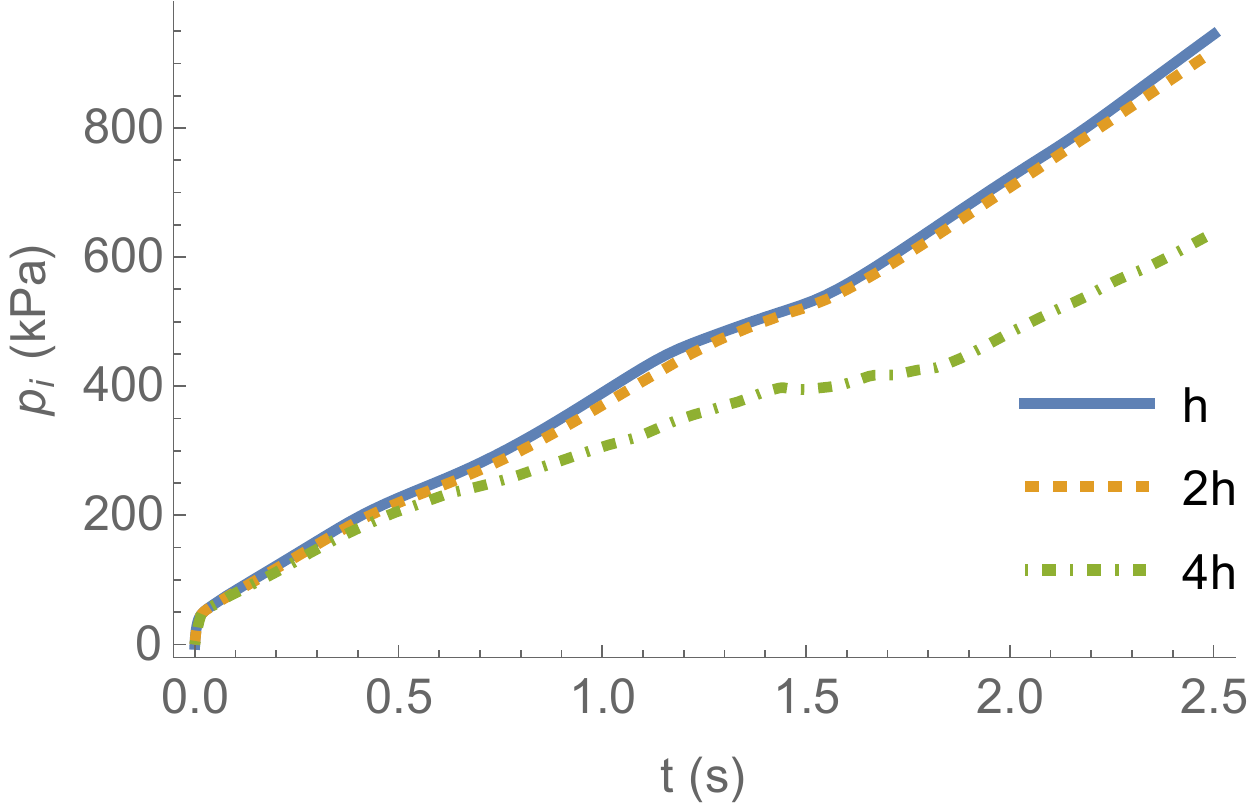}
	\end{overpic}
	\hspace{0.01\linewidth}
	\begin{overpic}[width=0.48\linewidth]{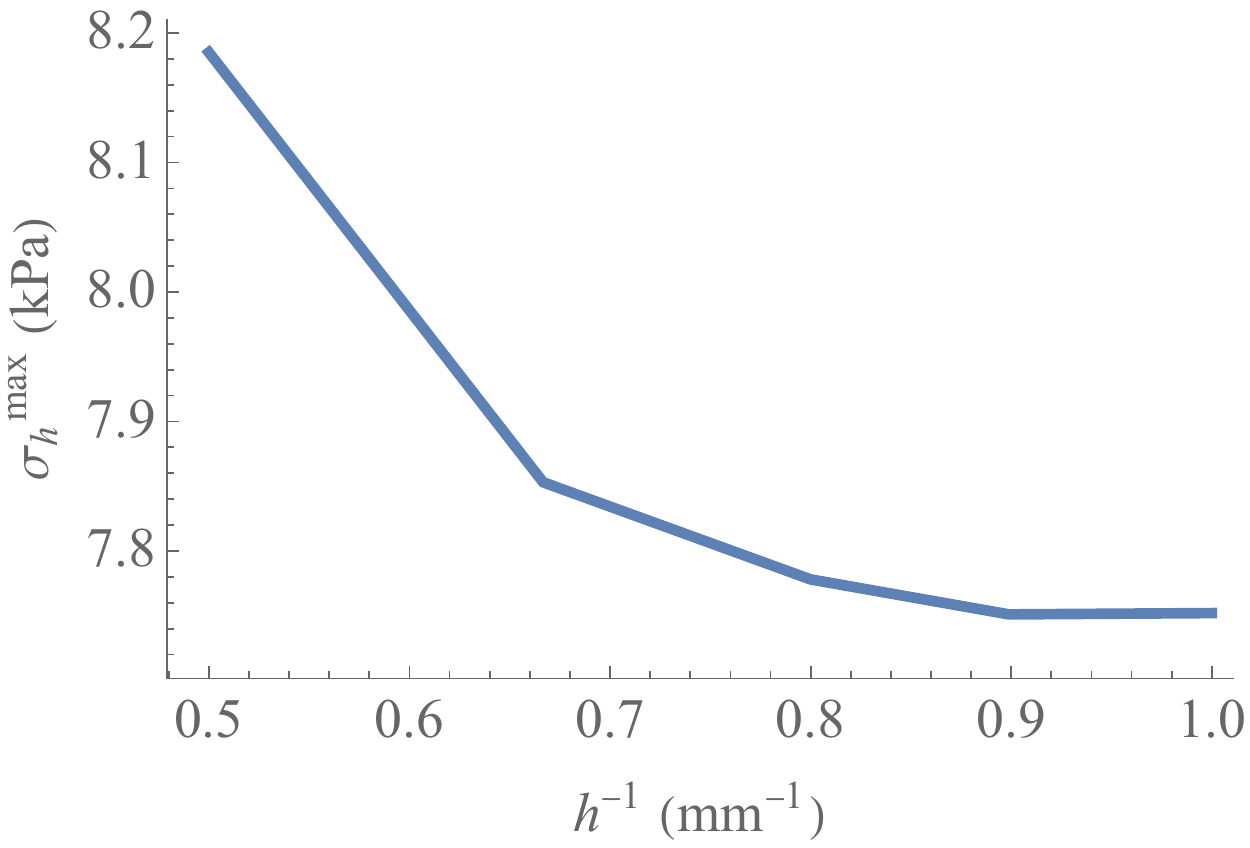}
	\end{overpic}
	\caption{Graphs of the injection pressure over time and of the maximum hoop stress for different mesh refinements, with reference element size $h=\SI{1.0}{\milli\metre}$.}
	\label{fig:mesh-conv}
\end{figure}

\subsection{Reproduction of a given experiment}\label{sec:reproduction}

The extent to which the model can reproduce qualitative aspects of a given Hele--Shaw experiment is now examined.  In particular, consider an experiment in which the cement fraction is 1.2\%, the oil viscosity is $\SI{300}{kcSt}$ (kilo centistokes) or \SI{0.29}{kPa.s}, and the injection flow rate is \SI{100}{mL/min}. 
In line with the provided experimental parameters, the model-based simulation parameters for this problem are summarized in Table~\ref{tab:table_summary} in the Appendix.  The random field shown in Figure~\ref{fig:ini-poro-fields}a is used as an initial condition for the porosity. 

The Hele--Shaw cell is represented in the simulations by an annulus of inner radius $r_i=\SI{5}{\milli\metre}$ and outer radius $r_o=\SI{53}{\milli\metre}$, as indicated in Figure~\ref{fig:BVP}.  
The inner radius is estimated by measuring the average radius of the inner cavity delimited by the beads in the experiment. Within the experimental estimated range of values for the initial average permeability, a value toward the lower end of this range is found to work best in the simulations, namely \SI{0.0015}{\milli\metre^2}. The elastic moduli of the porous skeleton are estimated from experimental measurements~\cite{MengLiJuanes-personal-2022}.

The simulation results are obtained with fracture parameters that scales with the cement volume ratio $C$ as follows.  For the fracture toughness, the experimentally-determined relationship \eqref{eq:Gc} is adopted. The nucleation energy $\psi_c $ governs damage initiation.  Assuming that crack nucleation occurs at a critical value $p_c$ of the injection pressure, a simple estimate of the critical fracture energy is given by (see equation 24 in \cite{Geelen2019})
\begin{equation}
	\label{eq:psic}
\psi_c=\frac{p_c(C)^2}{2E}.
\end{equation}
From the experimental measurements, the critical injection pressure $p_c$ that corresponded to crack initiation appears to scale linearly with the cement volume ratio, such that $p_c = \alpha C$. A magnitude of $\alpha=\SI{1}{kPa}$ is found to provide a good match with the experimental observations for crack initiation as a function of cement volume ratio.  Finally, the damage viscosity $\beta$ in \eqref{eq:summary_equations}$_5$ is also set to be an increasing function of the cement ratio.  The precise relationship is given in the next section, when the simulation results are calibrated against experimental observations.

Figure~\ref{fig:comparison-simu-MIT-exp} provides snapshots of the experimental results at the start of the experiment and at $t=\SI{2.0}{s}$,  along with simulation results of the damage and porosity fields at the corresponding times.  All simulation results are shown in the reference configuration.  The time $t=\SI{2.0}{s}$ corresponds to the moment in the experiment when the injection pressure reaches a plateau of approximately $\SI{470}{\kPa}$ (see Figure \ref{fig:injection-pressure}). To facilitate  comparisons between the experimental and simulation results at $t=\SI{2.0}{s}$, the water front is indicated with white contour lines.  For the experiment, this contour was extracted manually based on the color of the fluid (blue water vs.\ gray oil).  The fractured surfaces are also sketched on the experimental image as blue contours.  In the simulation results, the white contour corresponds to the saturation level $S=0.5$. For reference, the full saturation field at the start of the simulation and at $t=\SI{2.0}{s}$ is shown in Figure~\ref{fig:saturation}.  

\begin{figure}[h!]
	\centering
    \hspace{-0.13\linewidth}
    \begin{overpic}[width=0.39\linewidth]{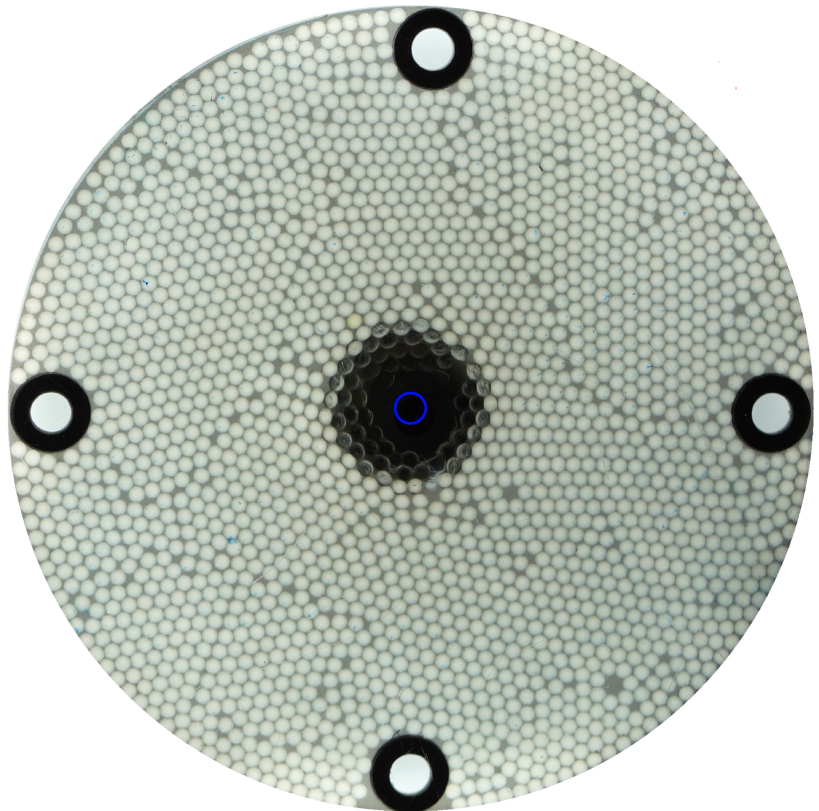}
		\put (-8,80) {\color{black} $a)$}
	\end{overpic}
	\hspace{0.09\linewidth}
	\begin{overpic}[width=0.39\linewidth]{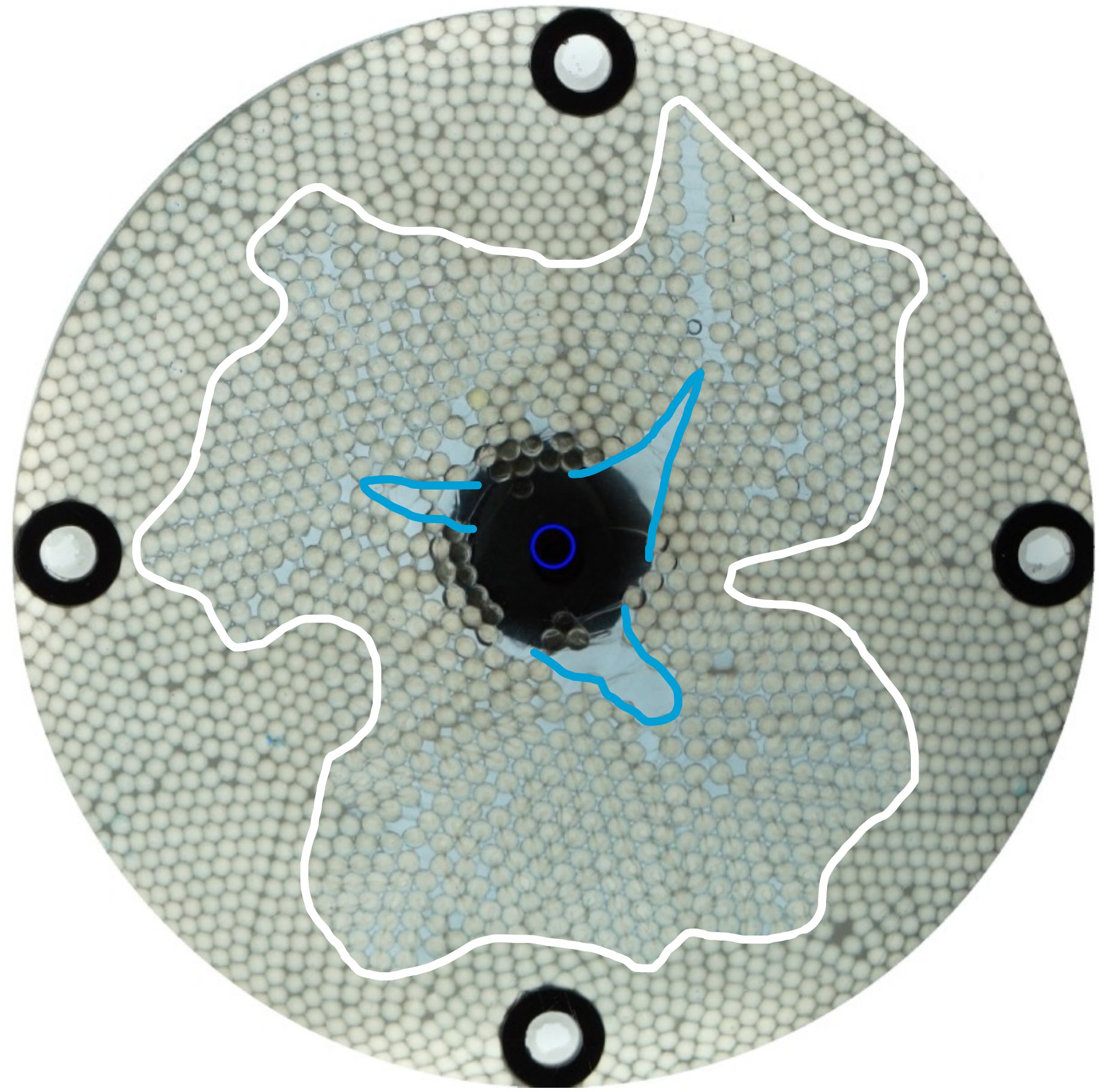}
		\put (-8,80) {\color{black} $b)$}
	\end{overpic}
	\begin{overpic}[width=0.48\linewidth]{poro_t=0}
		\put (-8,70) {\color{black} $c)$}
	\end{overpic}
	\hspace{0.01\linewidth}
	\begin{overpic}[width=0.48\linewidth]{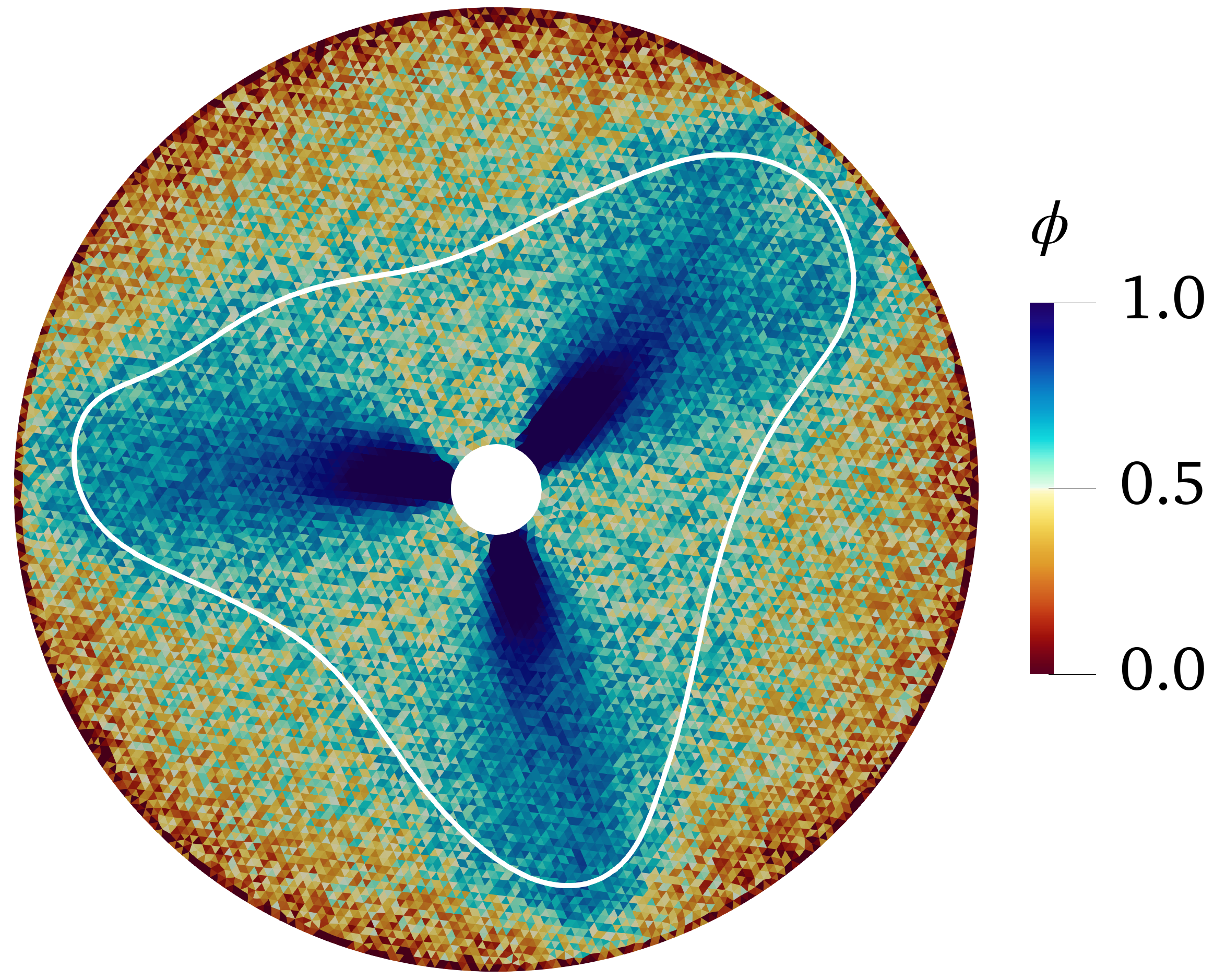}
		\put (-8,70) {\color{black} $d)$}
	\end{overpic}
	\begin{overpic}[width=0.48\linewidth]{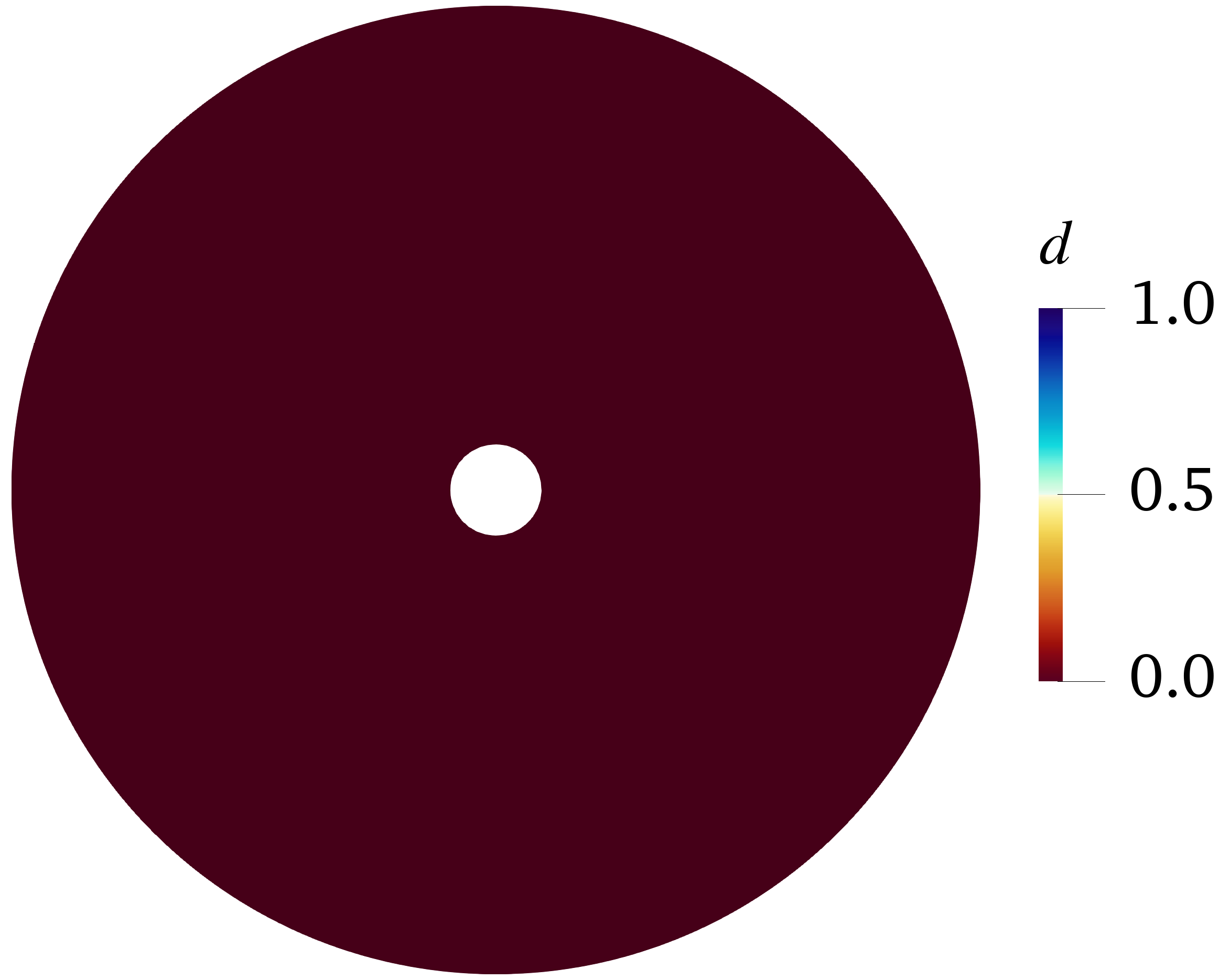}
		\put (-8,70) {\color{black} $e)$}
	\end{overpic}
	\hspace{0.01\linewidth}
	\begin{overpic}[width=0.48\linewidth]{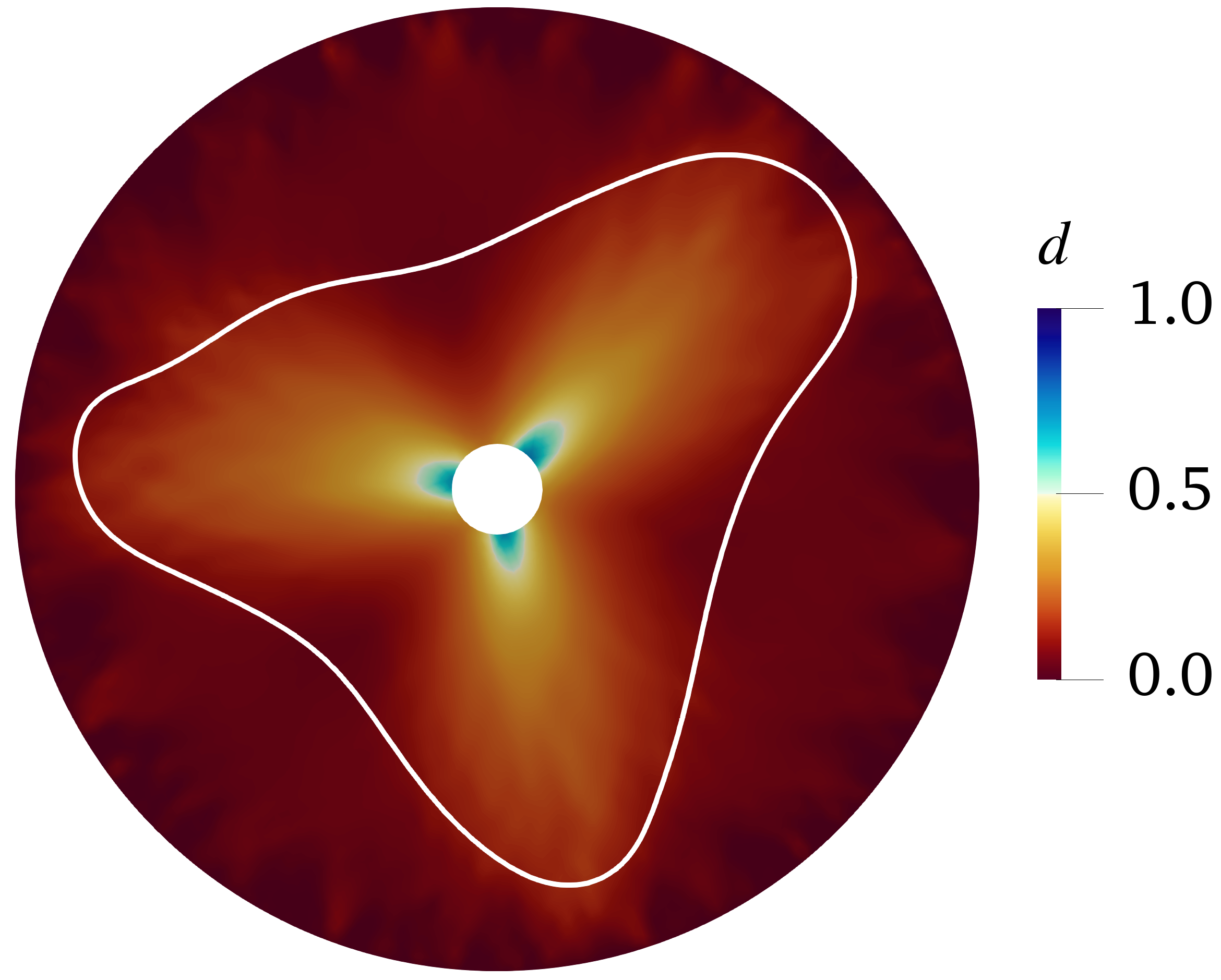}
		\put (-8,70) {\color{black} $f)$}
	\end{overpic}
	\caption{Comparison of the experimental results (a,b) with the simulation results (c, d, e, f). The left and right columns correspond to $t=0$ and $t=\SI{2.0}{s}$, respectively. The second row shows the porosity field, while the third row shows the damage field.  In the images on the right, the white contours indicate the leading edge of the invading fluid front.}
	\label{fig:comparison-simu-MIT-exp}
\end{figure}
\begin{figure}[h!]
	\centering
	\begin{overpic}[width=0.45\linewidth]{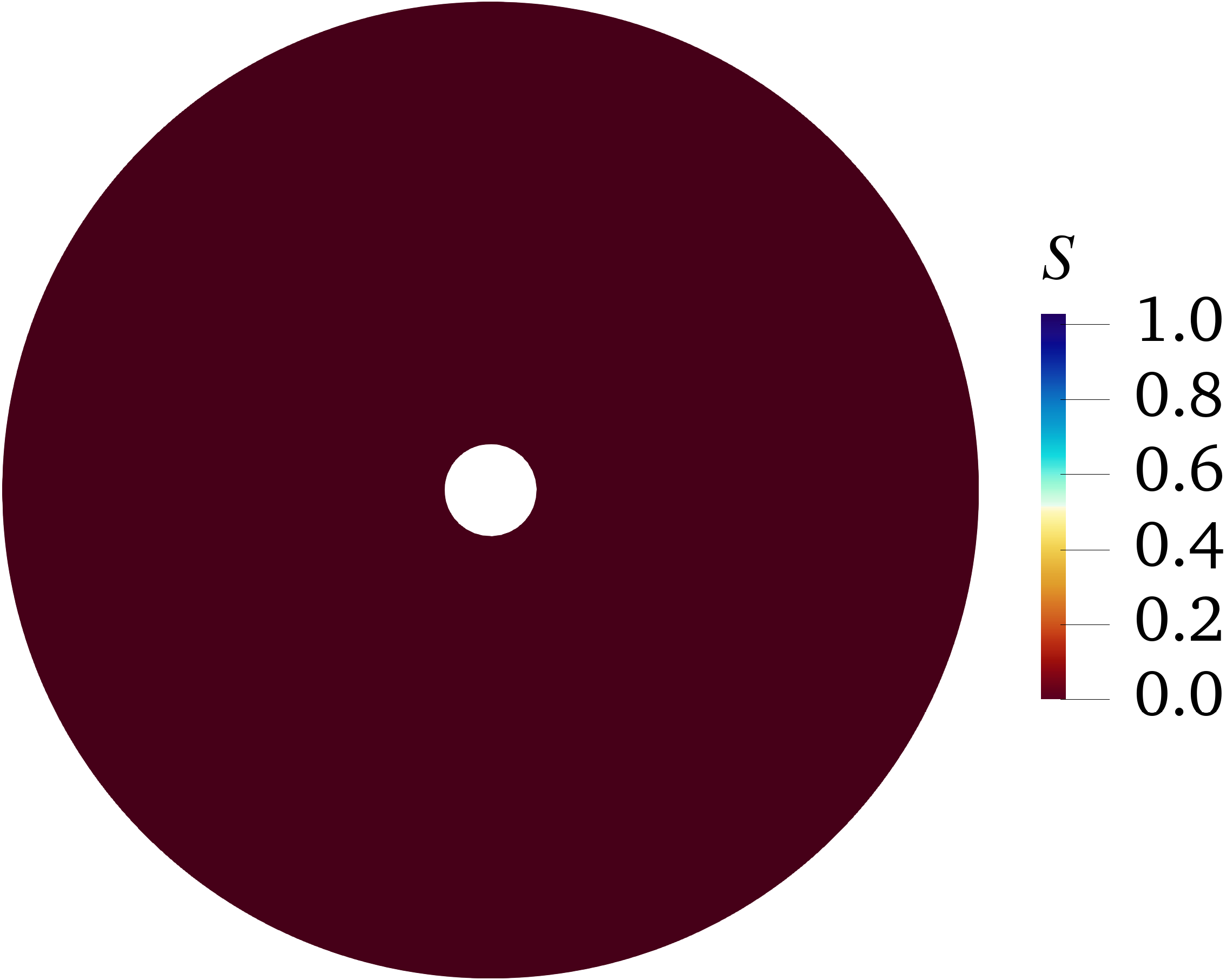}
		\put (-8,80) {\color{black} $a)$}
	\end{overpic}
	\hspace{0.05\linewidth}
	\begin{overpic}[width=0.45\linewidth]{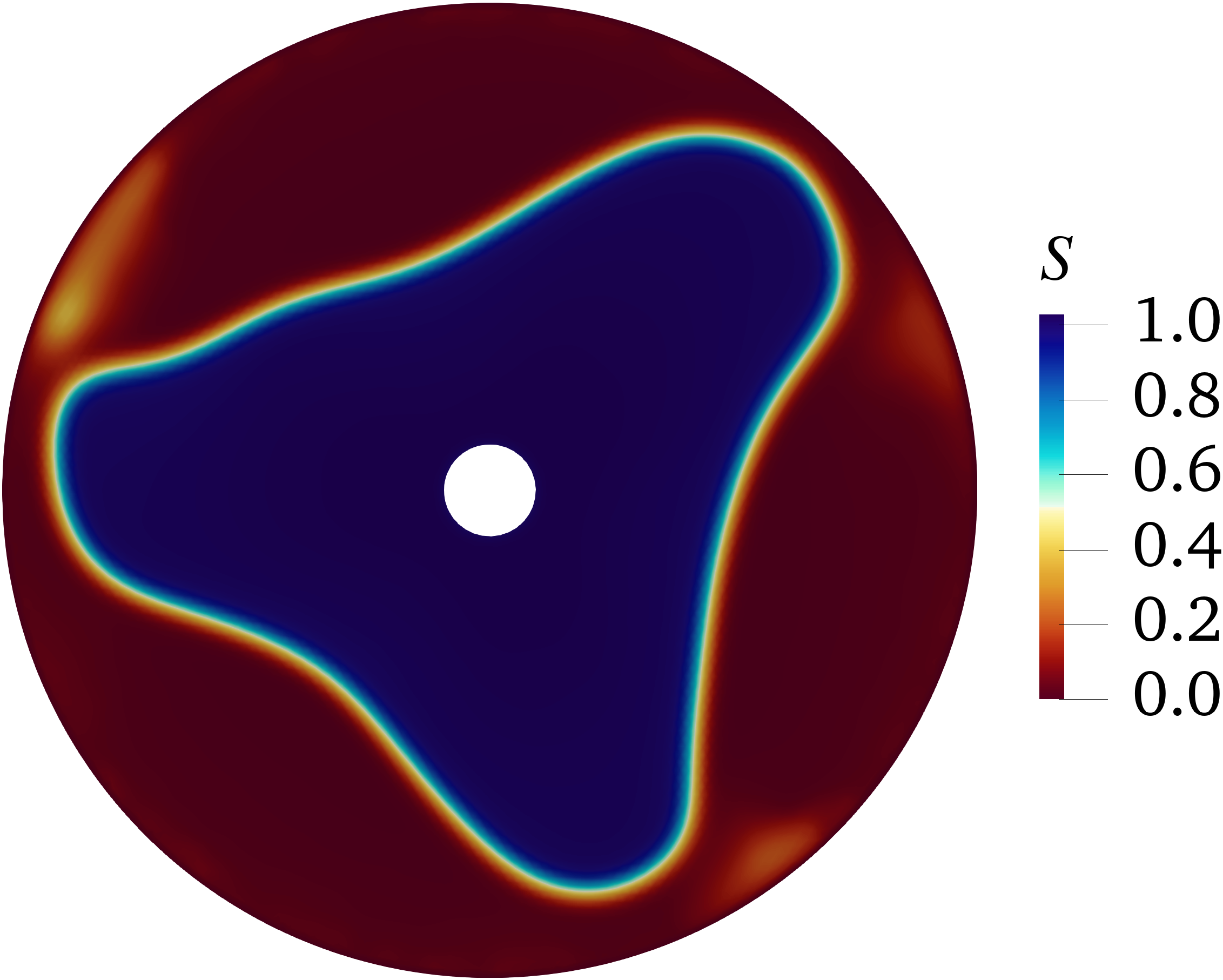}
		\put (-8,80) {\color{black} $b)$}
	\end{overpic}
	\caption{Saturation field corresponding to the white contour shown in Figure~\ref{fig:comparison-simu-MIT-exp} at $t=0$ (a) and $t=\SI{2.0}{s}$ (b).}
	\label{fig:saturation}
\end{figure}
An inspection of the results in Figure~\ref{fig:comparison-simu-MIT-exp} indicates that the model is able to replicate the main salient features of the water invasion. First, there are three main cracks, corresponding to regions of porosity equal to 1 in the simulations, preceded by partially damaged zones. At $t=\SI{2.0}{s}$, in the reference configuration, the cracks obtained in the simulations measure around \SI{11}{mm} on average, similar to the experiment. Secondly, at $t=\SI{2.0}{s}$, the saturation front from the simulation lies ahead of the cracks at a distance of \SI{45}{mm} from the center, calculated from an average of the distance of the three main invasion branches.  The analogous distance in the experiment is found to be approximately \SI{44}{mm}.  Both the experimental and simulation results indicate three types of fluid invasion: 1) into fully intact regions; 2) into partially open regions; and 3) into fully open regions (where the porosity $\phi= 1.0$).  Importantly, a fully open region (where $\phi=1$) does not necessarily coincide with a fully damaged region (where $d=1$). This is due to the damage viscosity term $\beta\dot{d}$.
In other words, fully open regions can still bear some tensile loading. This is justified by the presence of residual strands of glue observed in the experiment (see Figure~\ref{fig:comparison-simu-MIT-exp}b), which are assumed  sufficiently small so as not to impede flow but large enough to still allow the skeleton to transmit tractions. As a result, particularly in the present setting with viscous damage, it is important to interpret hydraulic cracks through the assessment of the porosity field and not the damage field. 

Figure~\ref{fig:injection-pressure} provides a comparison of the injection pressure as measured in the experiment to that extracted from the model-based simulation.  The results compare favorably, with a first stage of pressure linearly increasing with time, followed by a plateau starting around $t=\SI{1.7}{s}$.  In both the experiment and the simulation, the injection pressure peaks at approximately $\SI{470}{\kPa}$. The decrease of the rate of injection pressure coincides with the decrease of the rate of the average damage in the domain, as shown in Figure~\ref{fig:comparison_pressure_damage} in Appendix B. The delay between the damage plateau and the injection pressure plateau may be attributable to the damage viscosity.
\begin{figure}[h!]
	\centering
	\includegraphics[scale=0.7]{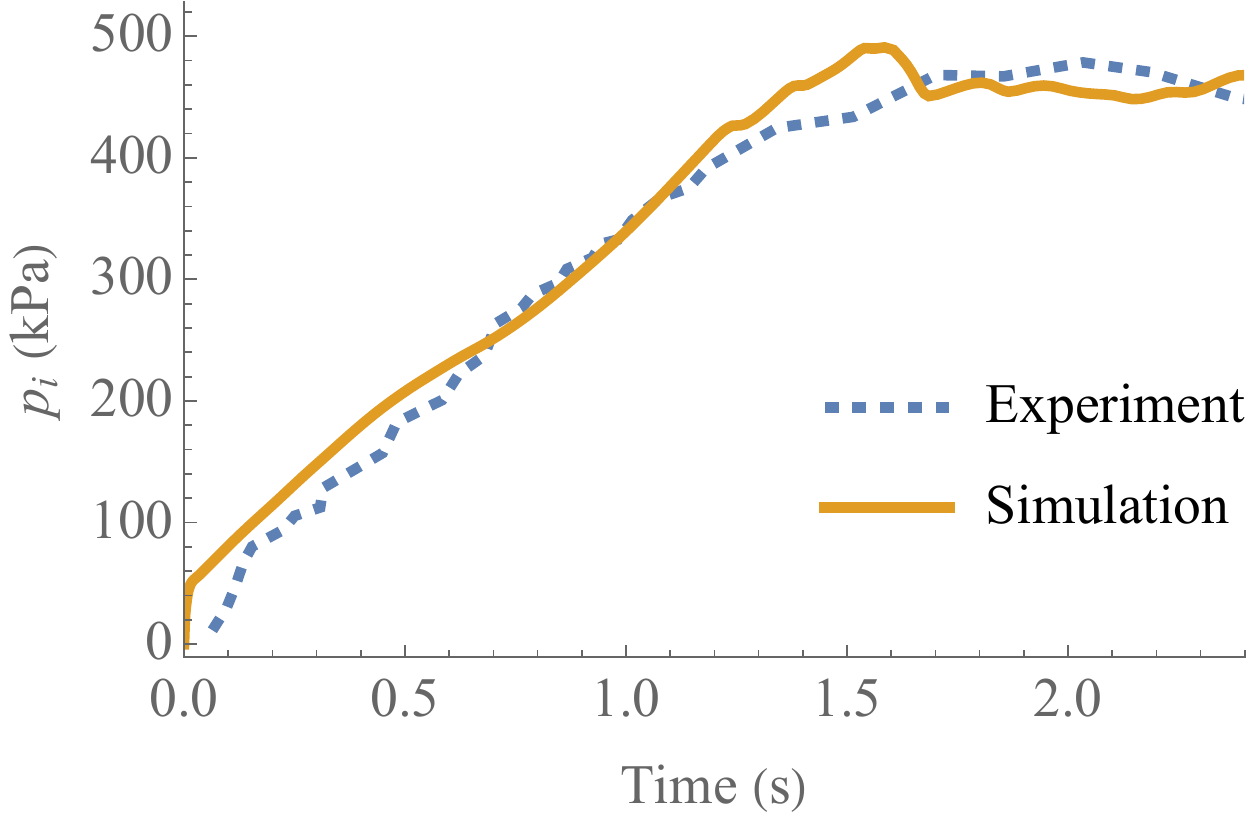}
	\caption[]{Injection pressure curve obtained from the experiment and matched in the simulation.}
	\label{fig:injection-pressure}
\end{figure}
Figure~\ref{fig:comparison-vol-strain} compares the volumetric strain response between the simulation and the experiment at $t=\SI{2.0}{s}$. In both cases, three regions of tensile volumetric strain are observed to develop out of the injection site. For both the simulation and the experimental results, these regions correspond to zones in which fracturing has occurred.  On one hand, the high tensile regions in the experiment are somewhat larger and more diffuse than those in the simulation.  On the other, the transition from regions of tensile volumetric strain to compressive volumetric strain is more abrupt in the experiment, at least in the areas between the three main branches.  Finally, the outermost boundary is seen to be in a state of compressive volumetric strain with a spatial variation that is more uniform in the experiment than the simulation.  This discrepancy may be due to the fact that the four outer clamps in the experiment are not explicitly modeled in the simulation. Alternatively, it could also be due to the fact that in the experiments the outer boundary containing the beads within the domain is not strictly rigid,  and therefore allows some release of elastic energy.
\begin{figure}[h!]
	\centering
	\begin{overpic}[width=0.38\linewidth, trim={0 -11cm 0 0}, clip]{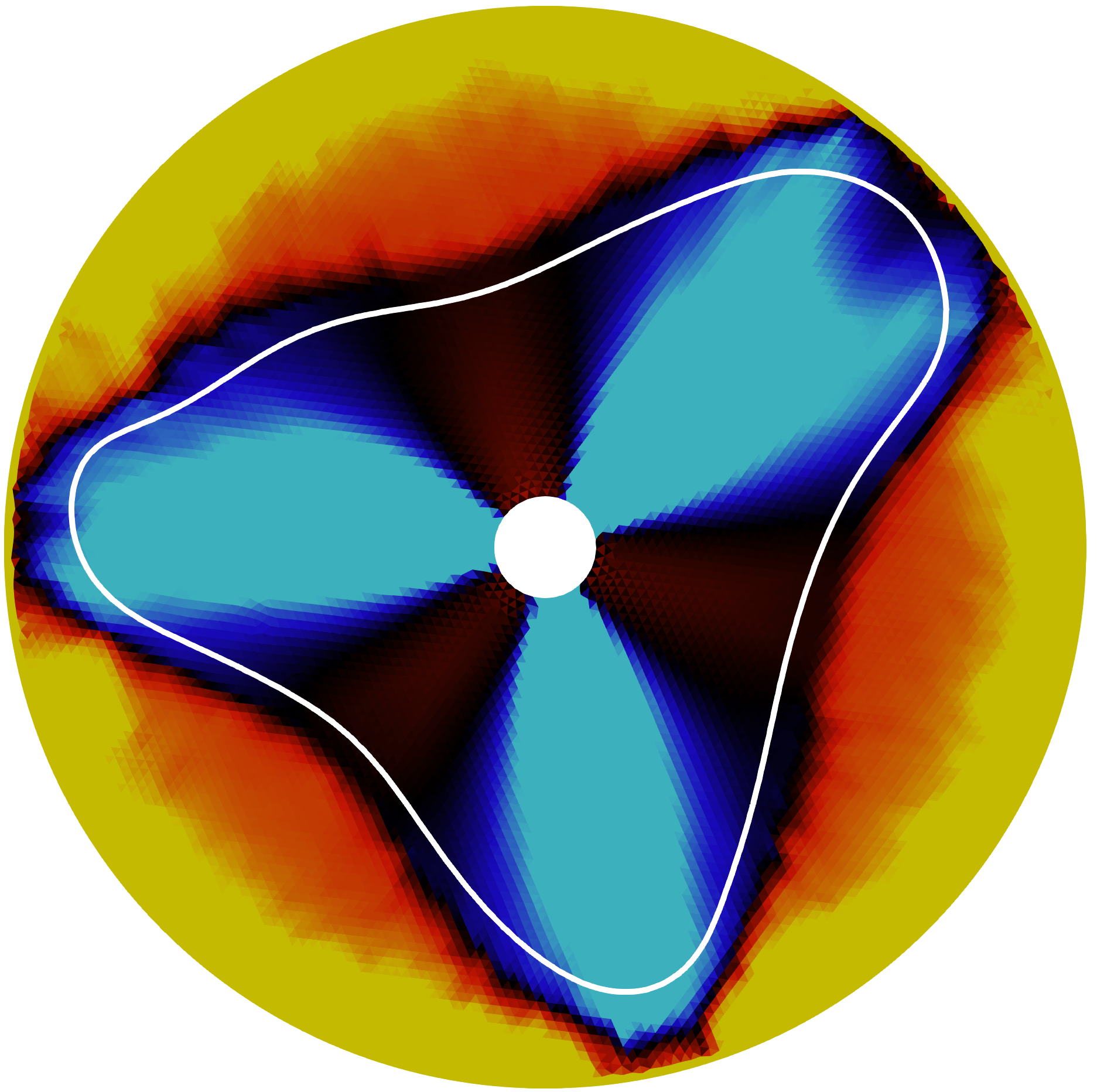}
	\end{overpic}
	\hspace{0.01\linewidth}
	\begin{overpic}[width=0.5\linewidth]{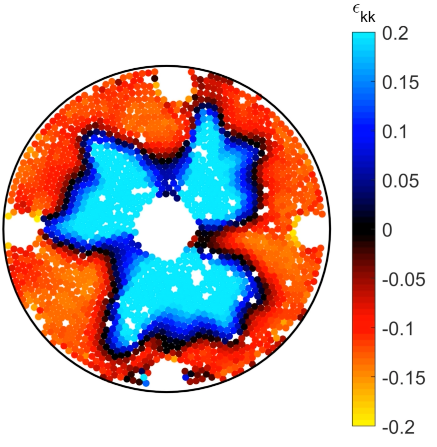}
	\end{overpic}
	\caption{Comparison of the volumetric strain field between the model-based simulation (left) and the experiment (right).  A fluid saturation of $S=0.5$ is indicated in the simulation result as a white contour. }
	\label{fig:comparison-vol-strain}
\end{figure}

Finally, Figure~\ref{fig:supplementary-fields} shows the simulation results for the pressure, hoop stress, permeability, and Darcy velocity of water at $t=\SI{2.0}{s}$.  To facilitate the interpretation of the results, a white contour line is once again used to denote the $S=0.5$ saturation front. The peak values of the pressure field can be seen to be concentrated within the invading fluid phase, as expected. The hoop stress is observed to be somewhat similar to the volumetric strain. The permeability is largest within the highly-damaged regions, until it reaches an upper bound of ($1.92^2/12=\SI{0.31}{\milli\meter^2}$).  Recall that this upper bound follows from assuming a Poiseuille flow confined by the two plates of the Hele--Shaw cell.  As anticipated,  the Darcy velocity is maximal in the damaged areas and at the water front. 
\begin{figure}[h!]
	\centering
	\begin{overpic}[width=0.4\linewidth]{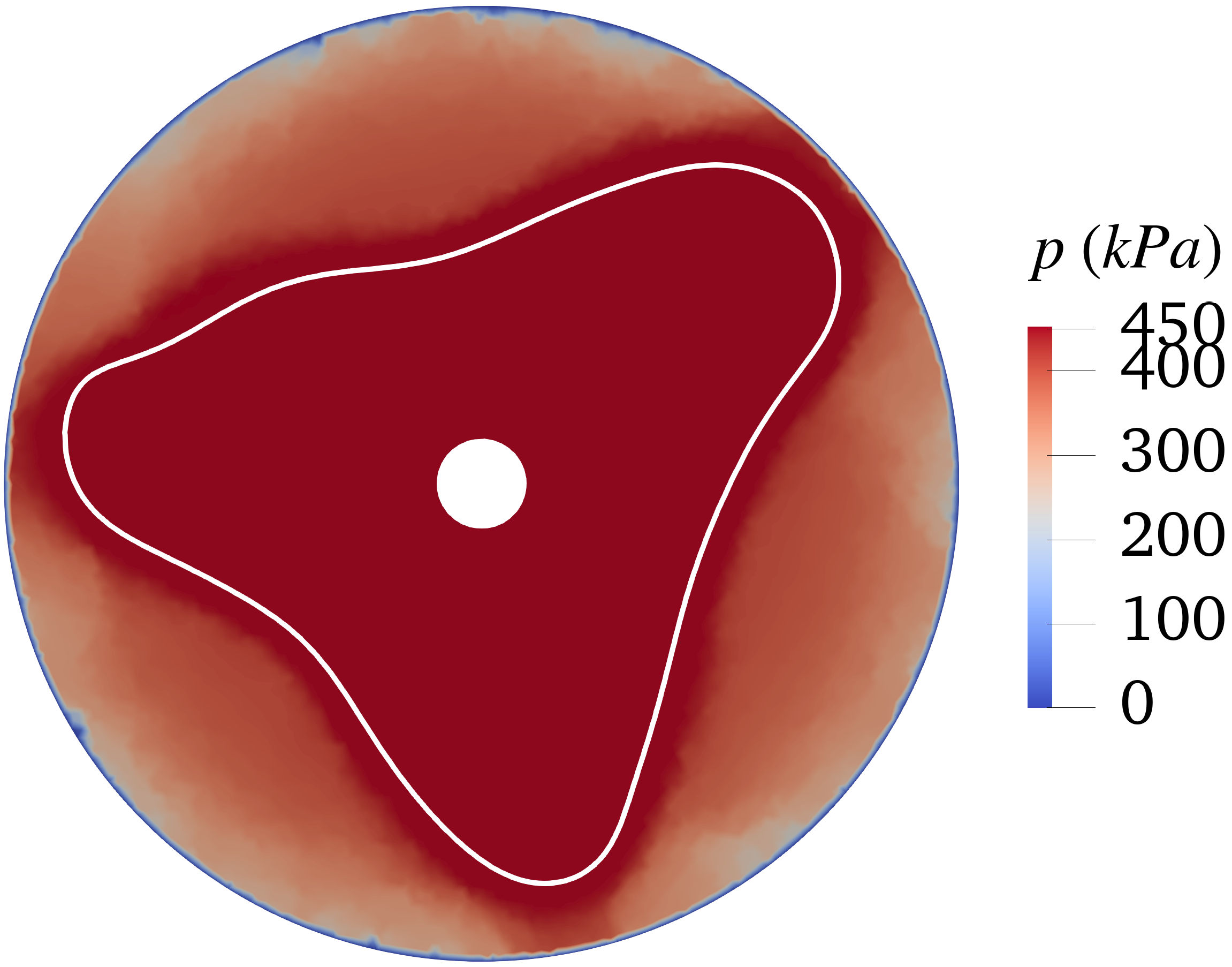}
		\put (-8,70) {\color{black} $a)$}
	\end{overpic}
	\hspace{0.1\linewidth}
	\begin{overpic}[width=0.4\linewidth]{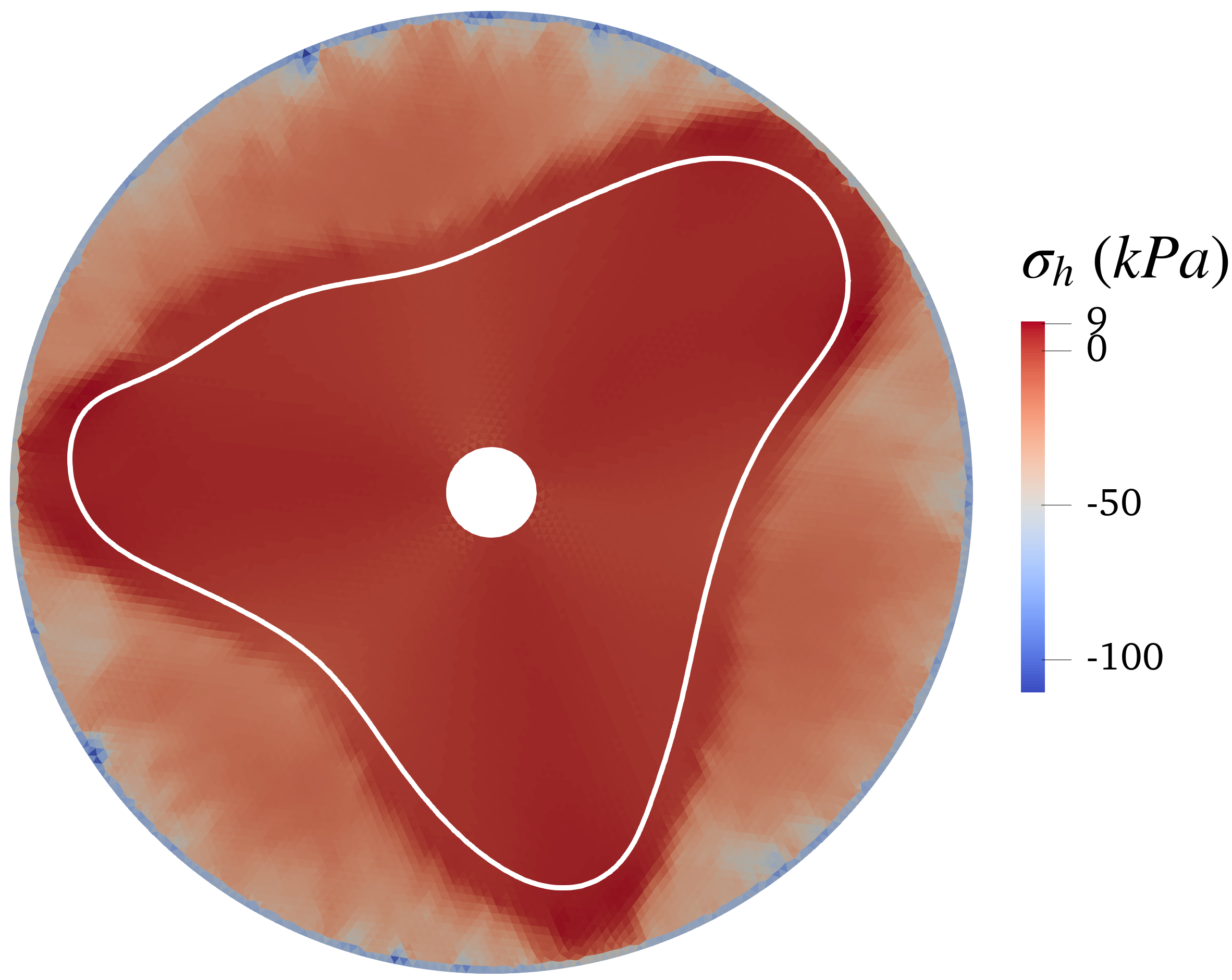}
		\put (-8,70) {\color{black} $b)$}
	\end{overpic}
	\begin{overpic}[width=0.4\linewidth]{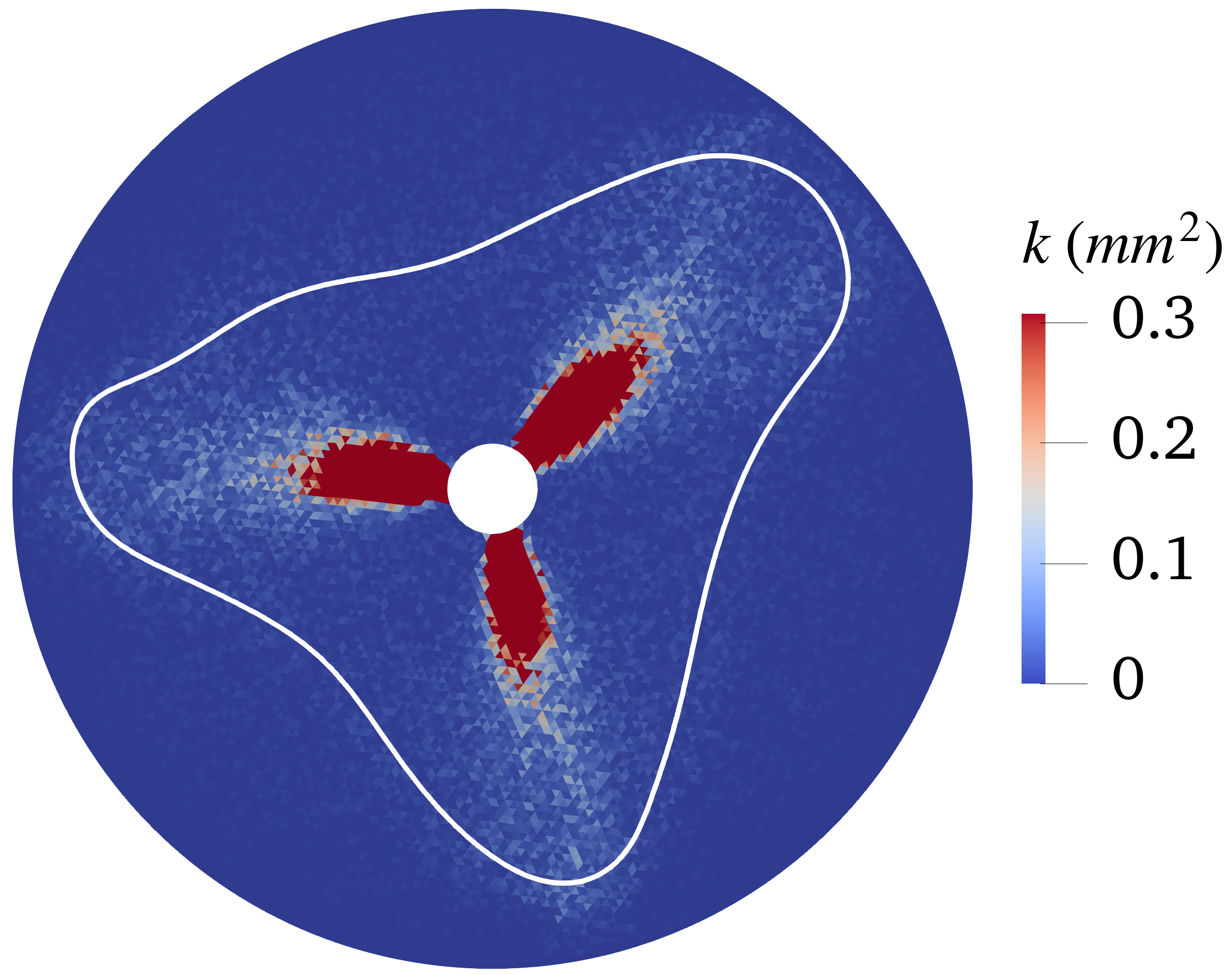}
		\put (-8,70) {\color{black} $c)$}
	\end{overpic}
	\hspace{0.1\linewidth}
	\begin{overpic}[width=0.41\linewidth]{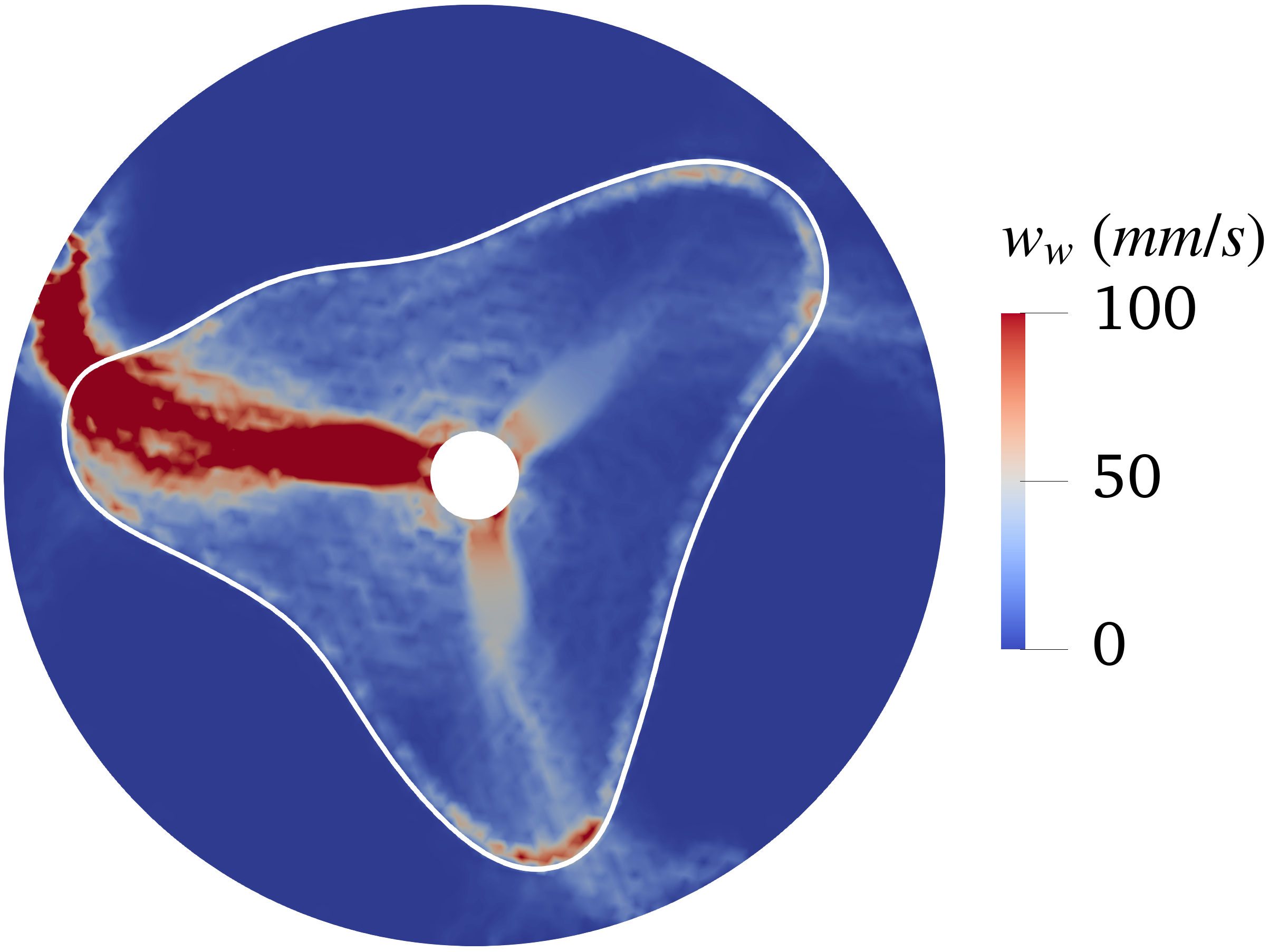}
		\put (-8,70) {\color{black} $d)$}
	\end{overpic}
	\caption{Simulation results for: a) the pressure; b) the hoop stress; c) the permeability; and d) the magnitude of the Darcy velocity of water.}
	\label{fig:supplementary-fields}
\end{figure}

\subsection{Reproduction of an experimental phase diagram}
\label{subsec:Reproduction of the experimental phase diagram capillary number VS cement fraction}

Experiments were conducted using monolayers of cemented beads with varying degrees of cement and a range of injection rates \cite{Meng2023}.  As expected, larger cement volume fractions gave rise to layers with increased fracture resistance. Figure~\ref{fig:Ca-VS-cement} shows a phase diagram delineating two regimes of response as a function of the modified capillary number $\text{Ca}^*$ and the cement volume ratio $C$.  The symbols in the diagram correspond to individual experiments or simulations, with the symbol type and color indicating the observed flow regime and source, respectively.   

Results obtained from simulations using the model developed in this work appear to replicate the phase diagram reasonably well, as indicated in Figure~\ref{fig:Ca-VS-cement}. The phase diagram can be interpreted as follows.  For a given cement ratio, fluid injection at relatively low capillary numbers gives rise to porous invasion without any fracturing of the monolayer.  As the capillary number is increased, the mechanical loading on the monolayer becomes large enough to give rise to fracturing.  Note that the transition from porous invasion to fracturing regime is progressive, in the sense that a simulation result labeled as porous invasion may display some small cracks, albeit with little influence on the invasion pattern.  Throughout this work, these two regimes are differentiated by measuring the length of the longest crack. If this length is larger (resp.\ smaller) than an arbitrarily small length, then the simulation result is labeled as belonging to the fracturing regime (resp.\ porous invasion regime). Following the discussion in Section \ref{sec:reproduction} regarding the importance of interpreting hydraulic cracks through the porosity field, crack lengths are calculated based on the distance from the center of the domain to the end of regions with porosity $\phi=1$.
The outcome of both experiments and simulations is that the threshold separating porous invasion from fracturing was found to scale exponentially with the cement volume ratio, as indicated by the solid black line in Figure~\ref{fig:Ca-VS-cement}.  

\begin{figure}[h!]
	\centering
	\includegraphics[scale=0.7]{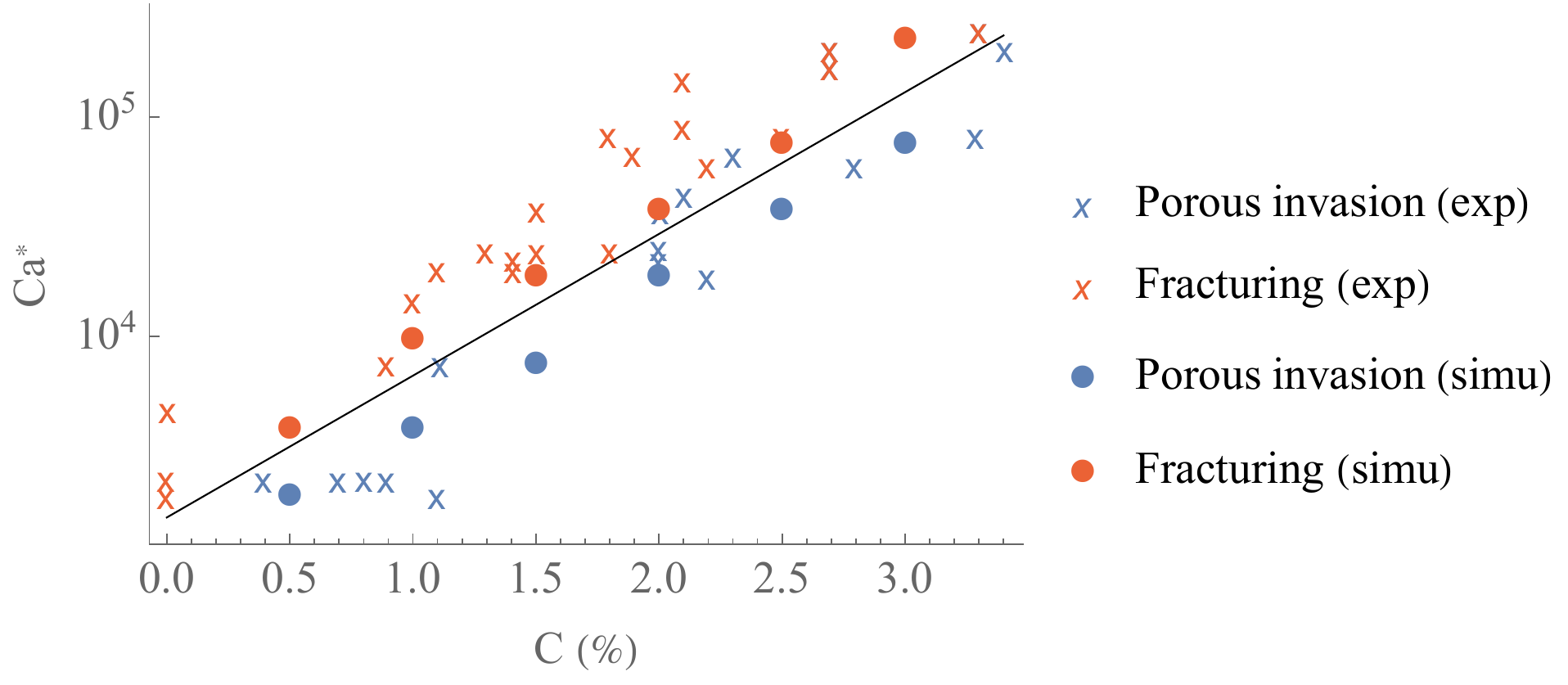}
	\caption[]{Phase diagram representing the modified capillary number $\text{Ca}^*$ vs.\  the cement volume ratio $C$. An exponential separatrix (black line) was found to separate a porous invasion regime from a fracturing regime for both the experimental and numerical results.}
	\label{fig:Ca-VS-cement}
\end{figure}

It bears emphasis that a key parameter in the model was adjusted to yield the best match between the experimental and simulation results in this phase diagram.  As mentioned in Section~\ref{sec:reproduction}, the damage viscosity $\beta$ is assumed to be an increasing function of the cement volume ratio $C$.  
 A relatively simple relationship that appears to provide a good match with the experimental phase diagram is $\beta = c \sqrt{C}$, with $c = \SI{65.7}{\kPa.s}$. 
Having sufficiently calibrated the model, in the remainder of this paper, simulations are employed to explore regions of the parameter space well outside the reach of the experimental setup.  

\subsection{Effect of the fracture parameters $\psi_c$, $G_c$, and $\beta$}

Before examining the influence of the main hydraulic parameters (permeability and viscosity), results illustrating the sensitivity of the model to the fracture parameters are reported.  The fracturing response of the system is governed by 3 parameters in the model: the nucleation energy $\psi_c$ (crack initiation), the fracture toughness $G_c$ (crack propagation), and the damage viscosity $\beta$ (rate effects). The sensitivity of the simulation results to variations in each of these parameters is illustrated in Figure~\ref{fig:effect-fracture-parameters}.  The middle column in the Figure corresponds to the base state that was used to make the comparisons in Section~\ref{sec:reproduction}.  For the sake of comparison, all simulation results are shown when the saturation front is at a similar distance from the outer boundary. 

\begin{figure}[h!]
	\centering
	\includegraphics[scale=0.75]{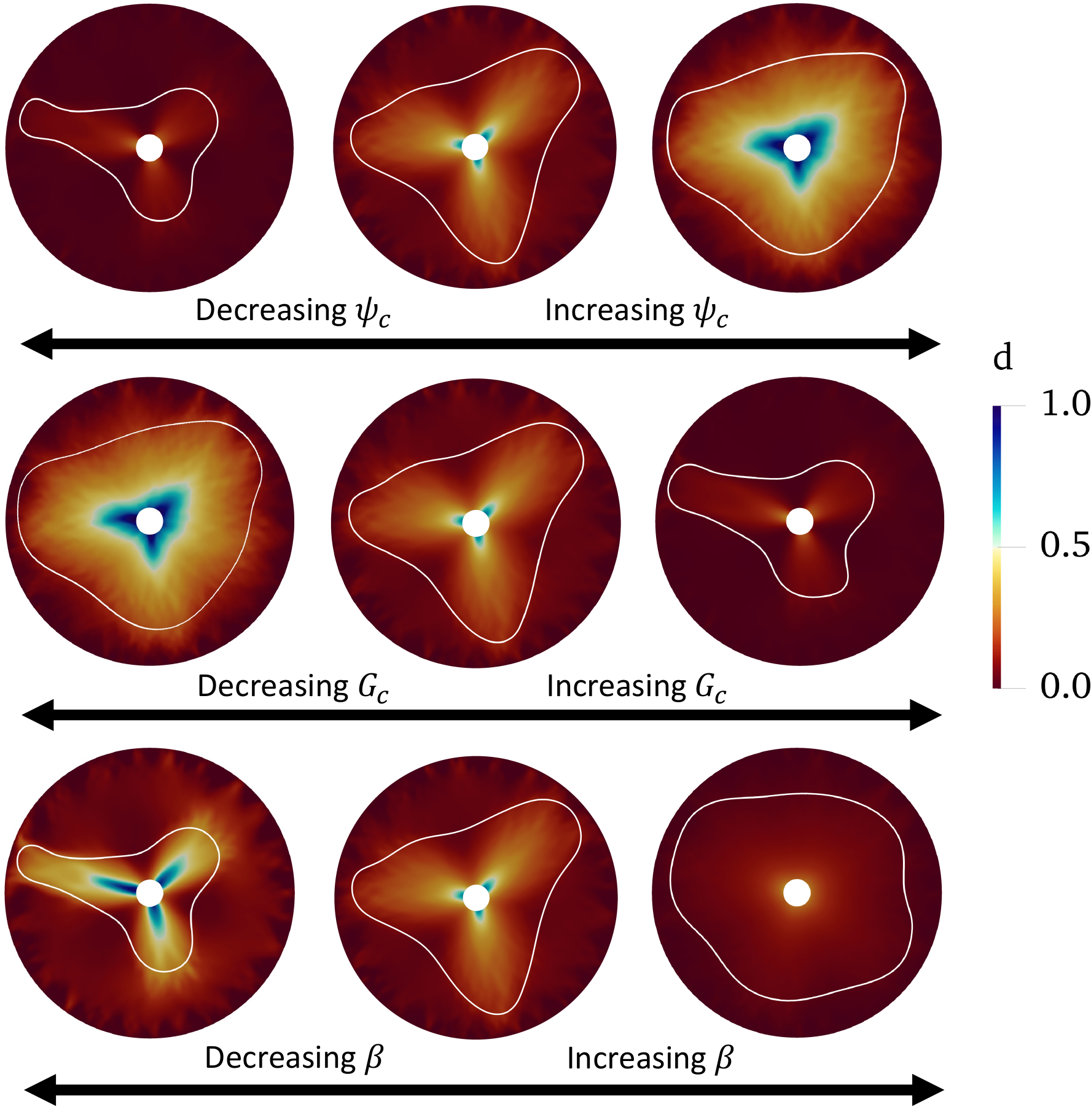}
	\caption[]{Effect of the fracture parameters $\psi_c$ (nucleation energy), $G_c$ (fracture toughness), $\beta$ (damage viscosity). The damage field is shown along with the white contour of the saturation front. The middle column corresponds to a reference simulation common to the three rows, obtained with the fracture parameters values stemming from the experimental study. The left and right columns are obtained by dividing and multiplying these reference values by 10, respectively. The simulations results are shown when the saturation front is at a same certain distance from the outer boundary.}
	\label{fig:effect-fracture-parameters}
\end{figure}

As expected, increasing either the fracture toughness $G_c$ or the damage viscosity $\beta$ is observed to inhibit crack propagation. Less expected is the effect of the nucleation energy $\psi_c$: while increasing its value delays crack initiation, the final cracks are observed to progress further in the domain. This may be explained by the accumulation of strain energy in the system while crack initiation is prevented until the energy reaches the higher threshold.

\subsection{Phase diagram representing fracturing number vs.\ branching number}

A wide range of the parameter space is now explored to infer the types of possible flow regimes predicted by the model. As described in the introduction, the model is expected to indicate various types of viscous flows, namely porous invasion, invasive and non-invasive fracturing, in possible combination with viscous fingering. The primary model parameters determining the type of flow are found to be, unsurprisingly, the 3 main fluid parameters $\eta_o$, $\eta_w$, and $k_0$, which are therefore widely varied in this section. Note that the characteristic permeability $k_0$ is taken as the initial average permeability in this section.

First, to determine whether the flow is of fracturing type or not, the fracturing number, denoted by $\text{N}_f$, is employed. Taking inspiration from Holtzman et al.~\cite{Holtzman2012}, the fracturing number is defined as the ratio of the driving force to the resisting force. The driving force is taken to be the viscous pressure drop $\delta p_{vis}$ introduced in \eqref{eq:viscous_pressure_drop}. The resisting force is the critical injection pressure $\delta p_{cri}$ at which the flow regime transitions from porous invasion to fracturing.  The fracture number thus reads:
\begin{equation}
	\label{eq:Nf}
\text{N}_f=\frac{\delta p_{vis}}{\delta p_{cri}}.
\end{equation}
While the driving force characterizes the invasion and contains only flow parameters, the resisting force characterizes the fracture resistance of the medium and depends on pertinent strength quantities such as the cement fraction. In this section, the fracture parameters are held fixed and, therefore, the value of $\delta p_{cri}$ is common for all simulations.

Second, to further characterize the type of flow within the fracturing and porous invasion regimes, a new dimensionless number is constructed, called the branching number and denoted by $\text{N}_b$, given by 
\begin{equation}
	\label{eq:Nb}
\text{N}_b=\frac{\eta_w l_0^2}{\eta_o k_0}.
\end{equation}
Note that the branching number is related to the Darcy number and the mobility ratio defined in \eqref{eq:dimensionless_groups} through $\text{N}_b=\text{M}^{-1}\text{Da}^{-1}$. 
For a fixed fracturing number $\text{N}_f$,  varying the branching number $\text{N}_b$ was found to give rise to two types of flow. In the porous invasion regime ($\text{N}_f<1$), low values of $\text{N}_b$ indicate a uniform porous invasion, whereas large values of $\text{N}_b$ indicate a viscous fingering instability. The number of branches of this instability was found to increase with $\text{N}_b$. In the fracturing regime ($\text{N}_f>1$), low values of $\text{N}_b$ indicate invasive fracturing, whereas large values of $\text{N}_b$ indicate non-invasive fracturing. The number of ``crack branches" was found to increase with $\text{N}_b$.

To illustrate these observations, Figure~\ref{fig:branching_number} shows the individual influence of the main parameters of $\text{N}_b$ on the type of flow. A first intuitive result is that decreasing the permeability tends to result in the invading fluid flow being more confined to the cracks, thereby indicating a transition from invasive to non-invasive fracturing.   Non-invasive fracturing is characterized by the invading fluid front being confined to the open channels that form. As for the influence of the fluid viscosities, similar trends are observed with increasing $\eta_w$ and $\eta_o$, as shown in the figure. Note that while $k_0$ and $\eta_w$ play a role that is consistent with the aforementioned rationale behind the branching number, it is not the case for $\eta_o$, at least individually. However, as a group, the branching number does play the role explained above, for fixed values of the fracturing number. Finally, the left simulation result of the third row of Figure~\ref{fig:branching_number} displays a new type of flow, which is the combination of the fracturing regime and viscous fingering. However, the branching number alone is not sufficient to identify the parameter space where this regime dominates; this will be done in combination with the fracturing number later in this section.

\begin{figure}[h!]
	\centering
	\includegraphics[scale=0.7]{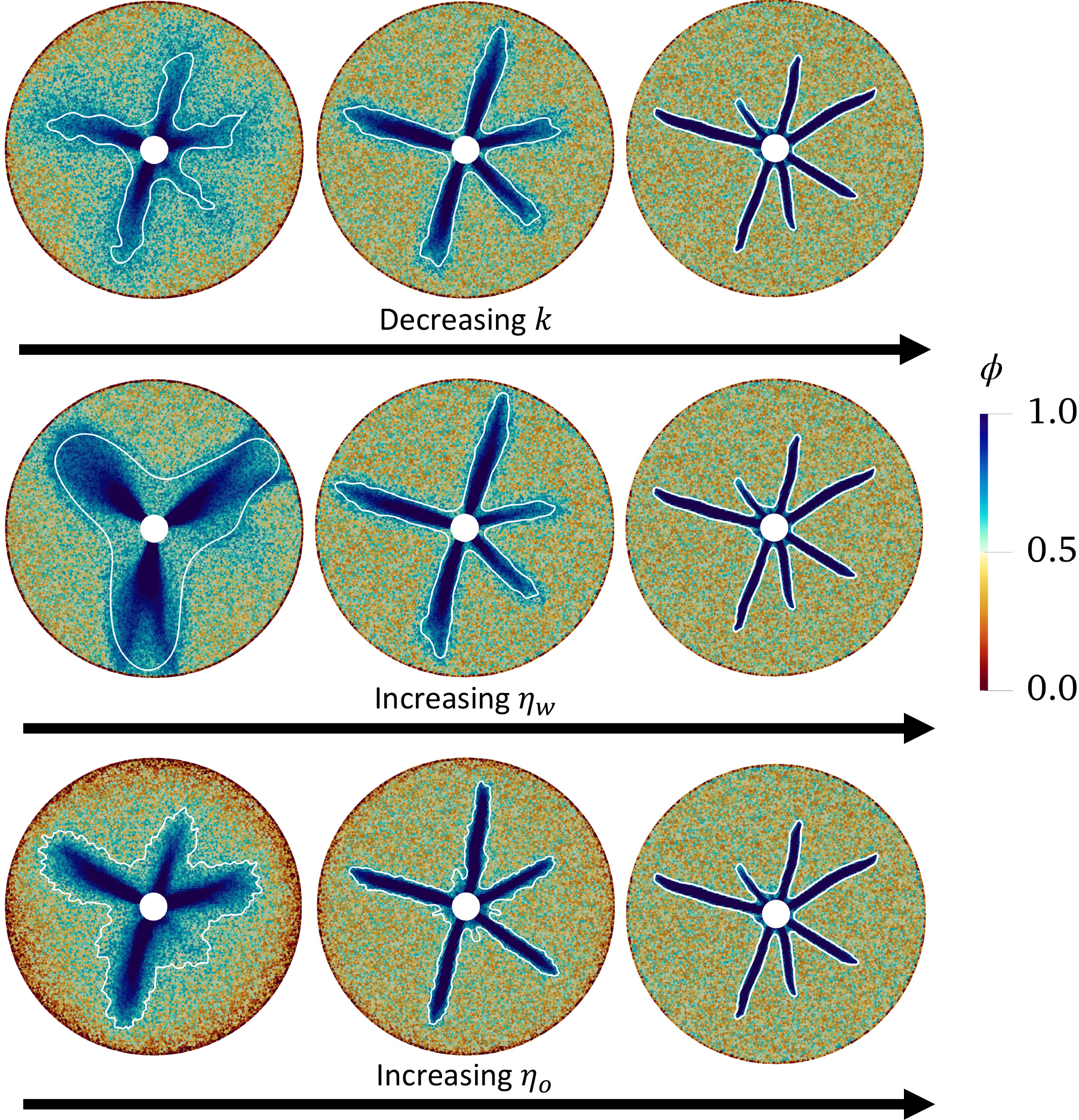}
	\caption[]{Plots of the porosity field and saturation front (white contour) illustrating the sensitivity to changes in various flow parameters.  Influence on the branching pattern of permeability (from left to right: $\bar{k}=\SI{5\times10^{-2}}{mm^2}$, $\bar{k}=\SI{10^{-2}}{mm^2}$, $\bar{k}=\SI{10^{-3}}{mm^2}$, at fixed $\eta_o=\SI{1}{kPa.s}$ and $\eta_w=\SI{10^{-3}}{kPa.s}$), of invading fluid viscosity $\eta_w$ (from left to right: $\eta_w=\SI{10^{-6}}{kPa.s}$, $\eta_w=\SI{10^{-4}}{kPa.s}$, $\eta_w=\SI{10^{-3}}{kPa.s}$, at fixed $\eta_o=\SI{1}{kPa.s}$ and $\bar{k}=\SI{10^{-3}}{mm^2}$), and of defending fluid viscosity (from left to right: $\eta_o=\SI{10^{-3}}{kPa.s}$, $\eta_o=\SI{10^{-2}}{kPa.s}$, $\eta_o=\SI{1}{kPa.s}$, at fixed $\eta_w=\SI{10^{-3}}{kPa.s}$ and $\bar{k}=\SI{10^{-3}}{mm^2}$).}
	\label{fig:branching_number}
\end{figure}

Following the previous qualitative study of the role of the main fluid parameters on the type of flow, a quantitative study is now presented, with the phase diagram shown in Figure~\ref{fig:diagram-perm-visco}.   
\begin{figure}[h!]
	\centering
	\includegraphics[scale=0.53]{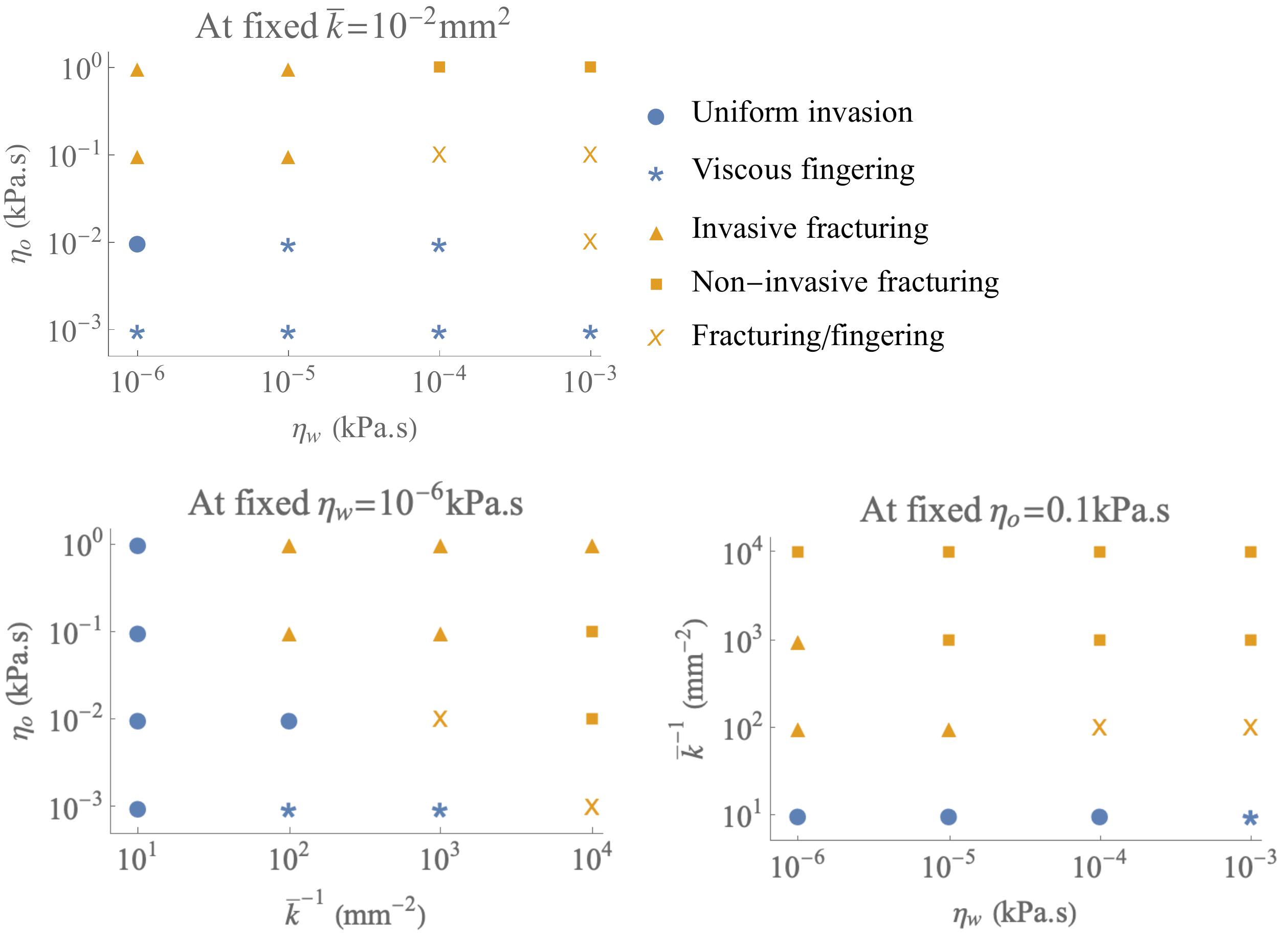}
	\caption[]{Phase diagram representing the effect of the three main flow parameters on the type of flow regime.}
	\label{fig:diagram-perm-visco}
\end{figure}
Note in particular that although the invading fluid viscosity $\eta_w$ has an effect on the branching pattern, it has a very limited influence on the fracture onset, as experimentally observed by Zhou et al.~\cite{Zhou2010}, and numerically confirmed by Carrillo and Bourg~\cite{Carrillo2021}.  The model-based simulations appear to recover this effect.  

To build a phase diagram of the fracturing number $\text{N}_f$ vs.\ the branching number $\text{N}_b$ representing all possible flow regimes, a large number of simulations are conducted over wide ranges of the parameters $\eta_o$, $\eta_w$, $k_0$. The resulting phase diagram is shown in Figure~\ref{fig:phase_diagram_Nf_vs_Nb}, with each symbol corresponding to an individual simulation and positioned at the corresponding values of $\text{N}_b$ and $\text{N}_f$.  For the resulting values of $\text{N}_b$, the characteristic length was chosen as the inner radius of the Hele--Shaw cell (i.e., \ $l_0=r_i=\SI{5}{\milli\metre}$).  

The particular flow regime indicated by each simulation is tagged as either a blue circle (uniform invasion), a blue star (viscous fingering), an orange triangle (invasive fracturing), an orange square (non-invasive fracturing), or an orange cross (fracturing/fingering).  The resulting regions where each type of flow dominates are also colored to help indicate threshold values for $\text{N}_f$ or $\text{N}_b$ that separate various regimes.  

Note that $\text{N}_b$ and $\text{N}_f$ are inherently different since, unlike the latter, the former is not a conventional dimensionless number since it is not the ratio of two competing forces.  Nevertheless, the phase diagram indicates that both the dimensionless numbers $\text{N}_b$ and $\text{N}_f$ do, in fact, play the role of regime delimiters, with values of $\text{N}_f=1$ and $\text{N}_b=1$ roughly separating fracturing vs.\ non-fracturing for the former, and uniform invasion vs.\ viscous fingering and invasive fracturing vs.\ non-invasive fracturing for the latter.
A new flow regime, mixing fracturing and fingering is found between the regions of non-invasive fracturing and viscous fingering, for $\text{N}_b >1$ and a certain range of $\text{N}_f$.
 Interestingly, in the region $\text{N}_b >1$, sufficiently large values of the fracturing number effectively prohibits viscous fingering, apparently favoring flow within the fractures over porous invasion.
 Finally, the red rectangle shown in  Figure~\ref{fig:phase_diagram_Nf_vs_Nb} represents the region of the diagram explored by the experiments. The latter was obviously able to showcase the two regimes of invasive fracturing and uniform invasion, while our simulations indicate the possibility of three more regimes that are beyond the range of conditions studied experimentally.
\begin{figure}[h!]
	\centering
	\includegraphics[scale=0.45]{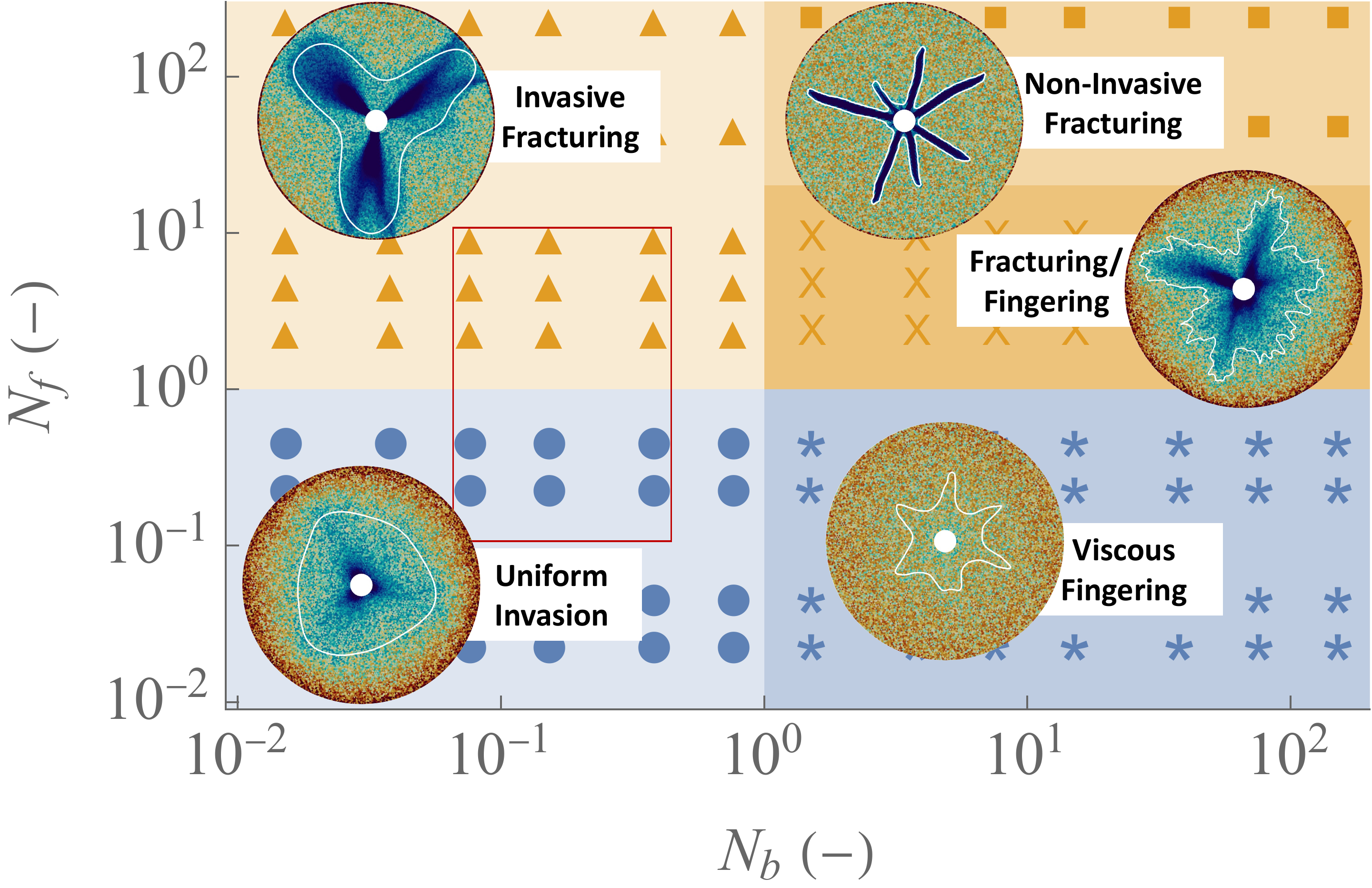}
	\caption[]{Phase diagram of the fracturing number $N_f$ vs the branching number $N_b$, mapping the five types of flow regimes. The red rectangle indicates the region of diagram explored by the experiments. The circular inserts are simulation results representing the porosity field along with the saturation front in white.}
	\label{fig:phase_diagram_Nf_vs_Nb}
\end{figure}

For practical purposes, this phase diagram can provide insight on the type of flow regime that can be expected at the field scale. For instance, it is well known that viscous fingering decreases the efficiency of fluid-fluid displacement in industrial applications (e.g., \ enhanced oil recovery \cite{Pinilla2021}). It is expected that industrial hydraulic fracturing may also suffer from this efficiency decrease, insofar as our results suggest the existence of a flow regime combining hydraulic fracturing and viscous fingering, provided that the branching number $\text{N}_b$ is sufficiently large. To estimate the latter, consider three types of field applications. First, in the context of oil recovery, if water ($\eta_w \sim \SI{10^{-6}}{kPa.s}$) is used to displace oil ($\eta_o\sim \SI{10^{-3}}{kPa.s}$), through a reservoir of permeability $k\sim \SI{10^{-9}}{mm^2}$, then $\text{N}_b\sim10^{10}$. The injection borehole is assumed to have a radius of $l_0\sim\SI{100}{mm} $. Second, in the context of CO$_2$ sequestration, where brine is displaced for water co-production, one can estimate $\text{N}_b\sim10^{12}$. The invading fluid, supercritical CO$_2$, is assumed to have a viscosity of $\SI{10^{-7}}{kPa.s}$, whereas the defending fluid, brine, is assumed to have the same viscosity as water, and the permeability is the same as in the first application. Third, in the context of geothermal reinjection, even though the invading fluid, water, is the same as the defending fluid, the viscosity of the former can be lower than that of the latter when its temperature is larger \cite{Mcdowell2016}. As in Macdowell et al.~\cite{Mcdowell2016}, when the viscosity of the invading and defending fluid is $\SI{10^{-4}}{kPa.s}$ and $\SI{10^{-3}}{kPa.s}$, respectively, and the permeability is $\SI{10^{-7}}{mm^2}$, then $\text{N}_b\sim10^{10}$. In all, the branching numbers for field applications are much larger than the range studied in Figure~\ref{fig:phase_diagram_Nf_vs_Nb}, so that the type of flow cannot be ascertained for the moment. It also bears emphasis that in real reservoirs, the flow regimes are not expected to be as distinguishable as in the simulations shown here. In particular, invasive fracturing, also called leak-off, can occur through existing natural cracks, even for tight geological formations \cite{Chen2021}. This means that invasive fracturing could occur for high values of $\text{N}_b$.

\section{Conclusion}
In this work, a double phase-field approach regularizing both cracks and fluid-fluid interfaces is introduced. Derived within the frameworks of continuum thermodynamics and linear poroelasticity, the model behavior is characterized by a tight three-way coupling between multiphase fluid flow, poromechanics, and fracturing. Through finite-element discretization and numerical simulations, the model was validated against Hele--Shaw experiments, both directly and via the reconstruction of a phase diagram discriminating porous invasion from hydraulic fracturing. The parameter space was then explored beyond experimental capabilities to discover a variety of flow regimes, including a combination of fracturing and viscous fingering.

The model employs many simplifying assumptions that permit relatively low computational cost while satisfyingly reproducing relatively complex experiments. However, in future works, it will be of interest to explore the influence of higher-fidelity modeling choices. First, the small strain assumption underpinning linear poroelasticity is at the limit of validity in the present setting and may be replaced by finite-strain kinematics, as in Paulin et al.~\cite{Paulin2022}. Second, interpreting the packing of the experimental beads as a continuum is at the limit of validity as well. This could be addressed by resorting to explicitly modeling the microstructure on one hand or by  experiments where the microstructural scale is much smaller than the domain size, on the other. Third, the binary fluid flow description can be enriched by replacing our Darcy--Poisseuille flow model by a Darcy--Stokes flow model, as introduced by Ehlers and Luo~\cite{Ehlers2017}, and Wilson and Landis~\cite{Wilson2016}. Fourth, provided light modifications such as the introduction of a capillary pressure, our model can accommodate for capillary flows in addition to viscous flows.

Ultimately, the model presented in this contribution could be used to predict and control the type of flows in field applications, as it bridges laboratory and field scales.   

\section*{Acknowledgements}

This work was the result of a collaboration between researchers at Duke and MIT, largely supported by a set of collaborative NSF research grants, in particular NSF grant CMMI-1933367 to Duke University, NSF grant CMMI-1933416 to MIT, and NSF grant CMMI-1826221 to Penn State. The support is gratefully 
 acknowledged. Any opinions, findings, and conclusions
or recommendations expressed in this material are those of the authors and
do not necessarily reflect the views of the National Science Foundation.    

\newpage
\appendix
\section{Summary table of the experiment's and model's parameters}
\begin{table}[h!]
	\centering
	\caption[]{Summary of the values chosen for the different model parameters.}
	\begin{tabular}{@{}ccccc@{}}
		\toprule
		\textbf{Name}  & \textbf{Symbol} & \textbf{Unit} & \textbf{Exp. value} & \textbf{Num. value}   \\ \midrule
		 \begin{tabular}[c]{@{}c@{}}height of\\ Hele--Shaw cell\end{tabular} & $h$ & \SI{}{\milli\metre} & 1.96 & / \\ \midrule
		 \begin{tabular}[c]{@{}c@{}}outer radius of\\ Hele--Shaw cell\end{tabular} & $r_o$ & \SI{}{\milli\metre} & 53 & 53 \\ \midrule
		beads diameter & $d$ & \SI{}{\milli\metre} & 2 & /  \\ \midrule
		packing fraction & $1-\phi_0$ & - & 0.60 & 0.60  \\ \midrule
		cement volume ratio & $C$ & \% &  $[0, 3]$ & /  \\ \midrule
  		oil viscosity & $\eta_o$ & \SI{}{\kPa.s} & $[0.029,0.29]$ & $[10^{-4}, 1]$ \\ \midrule
    water viscosity & $\eta_w$ & \SI{}{\kPa.s} & $10^{-6}$ & $[10^{-6}, 10^{-3}]$ \\ \midrule
		 permeability & $k$ & $\SI{}{\milli\metre^2}$ & $[(0.02d)^2,(0.06d)^2]$ & / \\ \midrule
		  permeability coeff & $\bar{k}$ & $\SI{}{\milli\metre^2}$ & / & $[10^{-4}, 10^{-1}]$ \\ \midrule
		surface tension & $\gamma$ & \SI{}{\kPa.\milli\metre} &  0.03 & 0.03 \\ \midrule
		\begin{tabular}[c]{@{}c@{}}Young's modulus\\ of granular pack\end{tabular} & $E$ & \SI{}{\kPa}  &  331 & 331 \\ \midrule
		\begin{tabular}[c]{@{}c@{}}drained bulk modulus\\ of granular pack\end{tabular} & $K$ & \SI{}{\kPa}  &  184 & 184 \\ \midrule
		\begin{tabular}[c]{@{}c@{}}Poisson ratio\\ of granular pack\end{tabular} & $\nu$ & \SI{}{\kPa}  &  0.2 & 0.2 \\ \midrule
		\begin{tabular}[c]{@{}c@{}}bulk modulus\\ of solid grains\end{tabular} & $K_s$ & \SI{}{\kPa}  &  1600 & 1600 \\ \midrule
		nucleation energy & $\psi_c$ & \SI{}{\kPa}  & / & $0.0015C^2$  \\ \midrule
		fracture toughness & $G_c$ & \SI{}{\kPa.\milli\metre}  & $(0.26C+1.00)^2$ & $(0.26C+1.00)^2 $ \\ \midrule
		damage viscosity & $\beta$ & \SI{}{\kPa.s}  & / & $65.7\sqrt{C}$  \\ \midrule
		 \begin{tabular}[c]{@{}c@{}} regularization length\\ for $S$\end{tabular} & $l_S$ & \SI{}{\milli\metre} & / &  4.0 \\ \midrule
	\begin{tabular}[c]{@{}c@{}} regularization length\\ for $d$\end{tabular} & $l_d$ & \SI{}{\milli\metre} & / & 4.0 \\ \midrule
   		 \begin{tabular}[c]{@{}c@{}} Inner boundary \\ condition for $\bfw_w$\end{tabular} & $w_i$ & \SI{}{\milli\metre/\second} & / &  50 \\ \midrule
  		 \begin{tabular}[c]{@{}c@{}} Inner boundary \\ condition for $\mu$\end{tabular} & $\mu_i$ & \SI{}{\kPa} & / &  0.006 \\ \midrule
 		 \begin{tabular}[c]{@{}c@{}} Outer boundary \\ condition for $\mu$\end{tabular} & $\mu_o$ & \SI{}{\kPa} & / &  0.02 \\ \bottomrule
	\end{tabular}
	\label{tab:table_summary}
\end{table}

\section{Comparison injection pressure and average damage}

\begin{figure}[h!]
	\centering
	\includegraphics[scale=0.9]{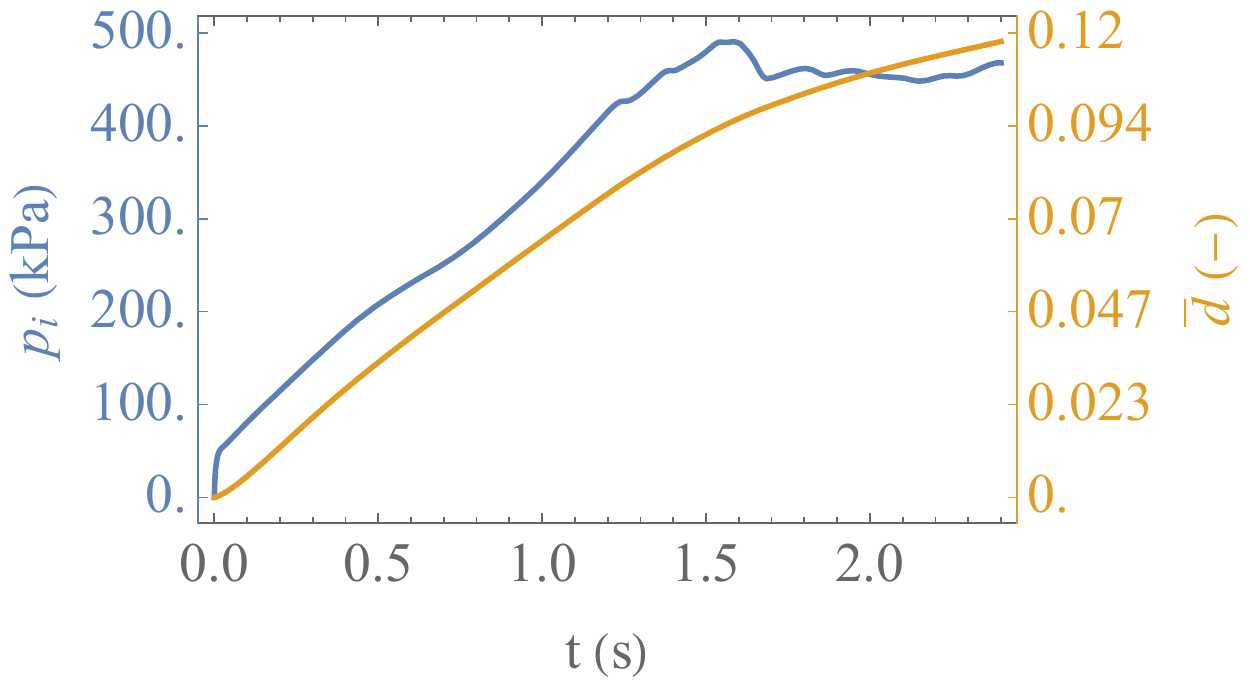}
	\caption[]{Comparison of the injection pressure curve with the average damage curve in the simulation reproducing the experiment.}
	\label{fig:comparison_pressure_damage}
\end{figure}

\clearpage
 \bibliographystyle{elsarticle_num} 
 \bibliography{Guevel_et_al_2022}

\end{document}